\documentclass[11pt,twocolumn]{article}
\usepackage{myusenix}

\usepackage{graphics}
\usepackage{url}
\usepackage{algorithm}
\usepackage[]{myalgorithmic}	
\usepackage{latexsym}		
\usepackage{amsthm}		
\usepackage{amssymb}		
\newtheorem{theorem}{Theorem}
\newtheorem{lemma}{Lemma}

\newcommand{\algname}[1]{\edef\algcs{\csname#1::name\endcsname}\textsc{\small\algcs}}

\newcommand{\algsig}[2]{\edef\algcs{\csname#1::name\endcsname}\edef\algcmd{\gdef\algcs{#2}}\algcmd\label{#1}\algname{#1}}

\newcommand{\algref}[1]{\algname{#1}}

\author{Petros Maniatis \hspace{4mm} Mary Baker\\ {\em Computer
Science Department, Stanford University}\\ {\em Stanford, CA 94305,
USA}\\
{\tt \{maniatis,mgbaker\}@cs.stanford.edu}\\
\url{http://identiscape.stanford.edu/}} 

\date{}
\begin{document}

\title{\bf Authenticated Append-only Skip Lists}

\maketitle

\thispagestyle{empty}

\begin{abstract}
\small In this work we describe, design and analyze the security of a
tamper-evident, append-only data structure for maintaining secure data
sequences in a loosely coupled distributed system where individual
system components may be mutually distrustful.  The resulting data
structure, called an Authenticated Append-Only Skip List (AASL), allows
its maintainers to produce one-way digests of the entire data sequence,
which they can publish to others as a commitment on the contents and
order of the sequence.  The maintainer can produce efficiently succinct
proofs that authenticate a particular datum in a particular position of
the data sequence against a published digest.  AASLs are secure against
tampering even by malicious data structure maintainers.  First, we show
that a maintainer cannot ``invent'' and authenticate data elements for
the AASL after he has committed to the structure.  Second, he cannot
equivocate by being able to prove conflicting facts about a particular
position of the data sequence.  This is the case even when the data
sequence grows with time and its maintainer publishes successive
commitments at times of his own choosing.

AASLs can be invaluable in reasoning about the integrity of system logs
maintained by untrusted components of a loosely-coupled distributed
system.
\end{abstract}

\section{\label{sec:introduction}Introduction}

Dependable systems rely heavily on logs of data, system events,
transactions, and security decisions.  Inspecting such logs while the
system is running (on-line) or after an exceptional event has caused the
system to cease operations (off-line) can help maintain accountability
through audit trails, repair failures through undo/redo logs, and
improve performance via profiling.

In distributed systems, especially those intended for loosely-coupled
communities of independent components (generally called
\emph{peer-to-peer systems}), it is frequently infeasible to maintain a
central system log; in fact, often there is no central authority that
can be trusted by all participating components to maintain a log
faithfully.  Instead, each component stores its own log of events
observed locally, or of interactions with other components.  The log for
the entire system does not exist physically; it is made up of log
fragments scattered around different system components.

These distributed log fragments must be perused for answers when a
failure occurs or a component is reported as misbehaving.  For example,
consider a distributed file system, where component $A$ requests that
component $B$ store datum $d$ and $B$ accepts.  Later $A$ attempts to
retrieve datum $d$ from $B$ and $B$ denies having this datum.  $A$ can
convincingly accuse $B$ of misbehavior only if it can show that, at
first, $B$ agreed to hold $d$ and, \emph{later}, $B$ denied having done
so.

In such a setting, logs can be a very sensitive and vulnerable system
resource.  A component, along with the entity that operates it, cannot
trust another component to retain the order of its locally logged events
or to refrain from changing events after it has logged them.  This
prevents $A$, in the example above, from using $B$'s log to justify its
accusation.  Similarly, an arbiter must be skeptical of an accusation
made by $A$ that relies on the integrity of $A$'s log.

The problem has been addressed in the literature via the use of
collision-resistant hash functions to link the contents of earlier log
entries to later
ones~\cite{Haber1991,Maniatis2002b,Schneier1998,Spreitzer1997}.  The
collision-resistance property of the hash functions used makes it very
difficult to ``rewrite history'' in a sensitive log, without causing
dramatic changes in the entire log.  However, with very few exceptions
(notably, work by Buldas et al.~\cite{Buldas1998}), no attention has
been paid to the scalability of such hash-based techniques when logs
grow very long and interesting sensitive log entries may have been
created years---and billions of log entries---ago.

In this paper, we analyze the security of the \emph{Authenticated
Append-only Skip List} (AASL).  The AASL is a novel data structure that
is designed for the efficient maintenance of and access to very large,
tamper-evident sequences of data.  AASLs provide a mechanism for
detecting structural corruption, such as modification, removal or
reordering of data, whenever those data are accessed.  We have used
AASLs in Timeweave~\cite{Maniatis2002b}, a mechanism that allows
components of a distributed system to maintain a local trustworthy view
of a global system log.

A distributed system component that maintains an AASL can compute
succinct one-way digests of the entire structure; these digests can
serve as a \emph{commitment} on the data structure contents and order,
and can be conveyed to other system components as such.  A remote
component wishing to establish whether a particular datum appears in
such a data sequence (membership) can request a \emph{proof} from the
maintainer of the AASL.  This proof can be verified against the digest
to which the maintainer has committed.

AASLs are guaranteed to prevent maintainers from proving conflicting
facts about a data sequence, even at different points in the evolution
of the sequence over time.  In this paper, we describe the construction
of AASLs and prove the security guarantees they offer.

\section{\label{sec:background}Background}

In this section, we describe related work that protects sensitive logs
from tampering, and related work on securing the contents of skip lists.

The integrity of public logs or commitment sequences has traditionally
been protected through the use of one-way hash functions.  Spreitzer et
al.~\cite{Spreitzer1997} describe how to protect the modification order
of a weakly consistent, replicated data system, by placing successive
write and read operations in a \emph{hash chain}; this is a linked list,
where every element is annotated with a label computed by hashing
together the value of the element and the label of the preceding
element.  Schneier and Kelsey~\cite{Schneier1998} propose a historic
integrity scheme for logs of untrusted or vulnerable machines.  Their
work protects access-controlled log entries against tampering or
unauthorized retroactive disclosure through hash chaining.  In secure
digital time stamping~\cite{Haber1991}, a digital notary places
documents in a hash chain, so as to be able to derive temporal
precedences between document commitments.

Unfortunately, reasoning about simple hash chains can be very expensive
when they grow long.  To check that a particular element occupies the
beginning of the chain, all the hashes between that element and the end
of the chain must be performed.  Buldas et al.~\cite{Buldas1998} improve
greatly on this linear cost; they describe optimally efficient hash
graphs that permit the extraction of such temporal precedences with
optimal proof sizes, on the order of the logarithm of the size of the
graph.

Goodrich et al.~\cite{Goodrich2001} retrofit skip lists for
tamper-evidence.  In that work, the authors propose an authenticated
skip list that relies on commutative hashing.  Anagnostopoulos et
al.~\cite{Anagnostopoulos2001} extend this construct to deal with
persistent data collections, where older versions of the skip list are
available, and they are each, by themselves, an authenticated skip list.
However, these structures are not designed to be append-only.  As a
result, they are not well-suited for tamper-evident logs: a malicious
maintainer can remove and then reinsert elements from the ``middle'' of
the structure across version changes.  A verifier must check vigilantly
that a log entry that interests him remains consistently in every new
version of the structure produced by the maintainer, which can be very
expensive when versions are produced frequently.

We have used the structure described in this paper in previous
work~\cite{Maniatis2002b} to preserve the historic integrity of a
loosely coupled distributed system.  Here we focus on a detailed design
and analysis of the security guarantees that the structure offers.

\section{\label{sec:skipLists:design}Design}

An Authenticated Append-only Skip List (AASL) is a data structure
conceptually based on skip lists~\cite{Pugh1990}.  Skip lists are sorted
linked lists with extra links, designed to allow fast lookup of the
stored data elements by taking ``shortcuts.''  The basic idea is to
enhance linked lists, which connect each element in the data sequence to
its successor, by also linking some elements to successors further down
the sequence.  Roughly half of the elements have links to their two-hop
successor, roughly a quarter of the elements have links to their
four-hop successor, and so on.  As a result, during traversal from
element $a$ to element $b$, the traversal path follows repeatedly the
longest available link from the current element that does not overshoot
the destination $b$, and thereby reaches $b$ in fewer steps than would
be possible by just traversing every intervening element between $a$ and
$b$.  Skip list traversals achieve logarithmic traversal path lengths in
the number of data elements in the structure, as opposed to the linear
paths offered by regular linked lists.

\subsection{\label{sec:skipLists:design:construction}AASL construction}

\begin{figure}
\centerline{\includegraphics{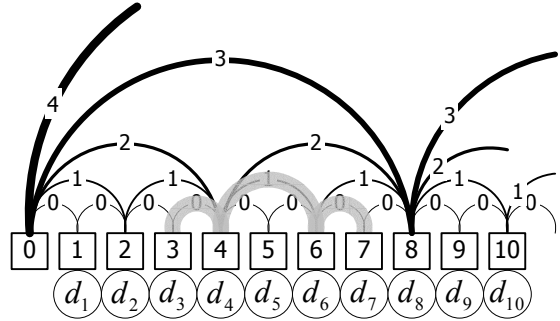}}
\caption[A deterministic skip list]{An example of a deterministic skip
list, containing 10 data elements.  Boxes denote element positions
(indices), and circles denote the actual element data.  We represent the
skip list as an overlapping set of linked lists, each at a different
level.  Pointers are marked with the level of the linked list to which
they belong.  The thick gray line outlines a traversal of the skip list,
from the 3rd to the 7th element.}
\label{fig:skipLists:AbstractSkipList}
\end{figure}

AASLs take advantage of the shortcut idea, described above, albeit in a
deterministic fashion as opposed to the randomized nature of the
original skip lists.  AASLs of $n$ elements consist of $\log_2 n$
coexisting linked lists, each designated by a different \emph{level}
number.  The linked list at level 0 is a regular linked list connecting
all elements in the data sequence.  The linked list at level 1 is a
linked list that only contains every other element from the original
data sequence.  The linked list at level $l$ contains every $2^l$-th
element of the original data sequence.  Element $i$ belongs to the
$l$-level linked list if and only if $i$ is divisible by $2^l$.
Figure~\ref{fig:skipLists:AbstractSkipList} illustrates this basic
structure.

AASL elements, in addition to their datum and their index number, carry
an \emph{authenticator}.  The authenticator $T^i$ for the $i$-th element
is a value derived via a few applications of a one-way hash function
$h$, such as SHA-1~\cite{SHA1}, to the datum $d_i$ of the $i$-th element
and the authenticators of the immediate predecessors of that element on
each of the linked lists in which it appears.

\begin{figure}
\centerline{\includegraphics{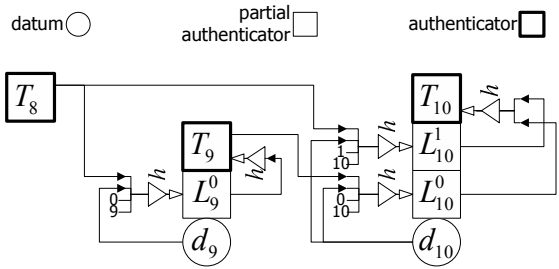}}
\caption[A detail of the skip list construction.]{An illustration of the
construction of the skip list authenticators for elements 9 and 10.  The
construction of the 8-th authenticator is not shown.}
\label{fig:skipLists:SkipListDetail}
\end{figure}

More specifically, an authenticator is computed in two steps (see
Figure~\ref{fig:skipLists:SkipListDetail}).  First, the \emph{partial
authenticators} for an element are computed, one for each list in which
that element participates.  A partial authenticator is a value computed
by hashing together the current index number, the current datum, the
current list level, and the authenticator of the preceding element on
that list.  The partial authenticator $L_i^l$ for the element in
position $i$ on the list at level $l$ is computed by
\begin{equation}
\label{eqn:skipLists:L}
L_i^l = h(i \| l \| d_i \| T^{i-2^l})
\end{equation}
where $\|$ denotes bit-string concatenation.  Second, the partial
authenticators are combined, again using the hash function, to produce
the element authenticator.  The authenticator $T^i$ of the $i$-th
element is computed by
\begin{equation}
\label{eqn:skipLists:T}
T^i=h(L_i^0 \| L_i^1 \| \ldots \| L_i^{f_i})
\end{equation}
where $f_i$ is the highest level of linked list in which the $i$-th
element appears.  $f_i$ is defined by the relation
\begin{equation}
\label{eqn:skipLists:f}
f_i = \{n : i = 2^n r \wedge \gcd(2,r) = 1\}
\end{equation}

A very useful property of skip lists is that they can be traversed from
a source element $i$ to a destination element $n$ ($i \geq n$) in a
number of steps that is logarithmic in the elements of the structure.
At every step, a linked list at the highest level is picked, among those
in which the current element participates, so as to travel the farthest
distance towards the destination, without overtaking it.
Algorithm~\ref{alg:skipLists:singleHopTraversalLevel} specifies how a
single hop is chosen for such a traversal.  The thick gray line in
Figure~\ref{fig:skipLists:AbstractSkipList} illustrates a traversal of
the structure.

\begin{algorithm}
\caption[Next-hop traversal level in AASL traversal.]{\small
\algsig{alg:skipLists:singleHopTraversalLevel}{SingleHopTraversalLevel}
$(i,n) \Rightarrow l$.  Return the highest linked list level $l$ that
must be followed in the AASL from element $i$ to element $n$, where $i
\geq n$.}
\begin{algorithmic}[1]
\small

\STATE $l \leftarrow 0$

\WHILE{$2^l$ divides $i$}

\IF{$i + 2^l \leq n$}

\STATE $L \leftarrow l$ \COMMENT{$L$-hop does not overtake $n$.}

\ELSE

\STATE Return $L$ \COMMENT{Last safe hop level.}

\ENDIF

\STATE $l \leftarrow l + 1$
\ENDWHILE

\STATE Return $L$ \COMMENT{The highest level possible for $i$.}

\end{algorithmic}
\end{algorithm}

\subsection{\label{sec:skipLists:design:membership}AASL Membership
Proofs}

The primary use of AASLs is to support authenticated answers to
membership questions, such as ``what is the 7-th element in the AASL?'',
while maintaining the append-only property of the AASL.  To accomplish
this functionality, it is important, first, that the party asking the
question (the \emph{verifier}) know in which AASL she is asking that
question; and, second, that once the verifier receives a response, she
holds that response as unequivocal for the AASL in question.

An AASL is uniquely determined by a \emph{digest}.  This is the
authenticator of the last appended element into the structure.  The
maintainer of an AASL conveys this short value to potential verifiers as
\emph{commitment} to the exact contents of the AASL.  A verifier who
receives such a digest verifies all subsequent exchanges with the
maintainer against this digest.

A response to a membership question on the contents of an AASL consists
of a \emph{membership claim} and a \emph{membership proof}.  A
membership claim has the form ``Data element $d$ occupies the $i$-th
position of the AASL whose $n$-th authenticator is known to the
verifier,'' and is denoted by $\langle i, n, d \rangle$.  The
corresponding membership proof is denoted by $E^{i, n, d}$.  This proof
convinces the verifier that, first, the maintainer had decided what the
$i$-th value $d$ would be before issuing the $n$-th authenticator;
second, the maintainer cannot authenticate any other value $d' \neq d$
as the value of the $i$-th element of the AASL with the known $n$-th
authenticator $T$.

The AASL maintainer constructs the membership proof $E^{i, n, d}$ by
traversing the AASL from the $i$-th to the $n$-th element, hop by hop,
as described by \algref{alg:skipLists:singleHopTraversalLevel}.  For
every encountered skip list element $j$, the maintainer constructs a
\emph{proof component} $C^j$ that consists of the $j$-th data element
and the authenticators of its predecessors on all the linked lists in
which it appears: $C^j = \langle d_j; \langle T^{j-2^l} : 0 \leq l \leq
f_j \rangle \rangle$.  The sequence of all proof components makes up the
membership proof $E^{i,n,d} = \langle C^j: j \in S^{i, n}\rangle$, where
$S^{i, n}$ is the sequence of elements traversed from $i$ to $n$.  The
appendix contains Algorithm~\ref{alg:skipLists:singleElementComponent},
which describes the construction process for a single proof component,
and Algorithm~\ref{alg:skipLists:constructWholeMembershipProof}, which
outlines the overall proof construction process.

The verifier processes a membership proof against the AASL authenticator
that it holds to verify the validity of a membership claim.  The
verification process mimics the proof construction process.  The
verifier's job, however, is to make sure that the purported proof is
well-formed and yields the known authenticator starting with the element
datum and position in the maintainer's membership claim.  The
verification may succeed with a positive result, which means that the
claim is true; it may succeed with a negative result, which means that
the claim is false, i.e., it \emph{cannot} be true; and it may fail, in
which case nothing is known about the claim, except that the supplied
proof is inappropriate for the given claim.

For every element $j$ in the traversal from the $i$-th to the $n$-th
element, the verifier checks that the corresponding component $C$ in the
proof is formed as component $C^j$ should be formed; he then uses that
component to compute what the $j$-th authenticator should be based on
that component.  Furthermore, since, during traversal, the authenticator
of a traversed element is always used in the computation of the
authenticator of the next traversed element, the verifier must check
that the authenticators it computes in earlier steps of the verification
process are consistent with those used in later verification steps.
Finally, the proof must be checked for applicability, that is, it should
match the claim it purportedly proves: if a membership proof claims to
prove the membership claim $\langle i, n, d \rangle$, then the datum in
the first proof component should be $d$.

Algorithm~\ref{alg:skipLists:processProofComponent} details how a single
proof component is handled by the verification process.
Algorithm~\ref{alg:skipLists:processWholeMembershipProof} details the
overall proof verification process, making use of the single-component
proof verification from
Algorithm~\ref{alg:skipLists:processProofComponent}.

\begin{algorithm}
\caption[Process a proof component in an
AASL.]{\small\algsig{alg:skipLists:processProofComponent}{ProcessProofComponent}
$(j, C) \Rightarrow T$.  Process the proof component $C$ that
corresponds to the $j$-th element in an AASL, and return the resulting
$j$-th AASL authenticator.}
\begin{algorithmic}[1]
\small

\STATE $\langle d; \langle T_0, T_1, \ldots, T_F \rangle \rangle
\leftarrow C$ \COMMENT{Parse $C$.}

\IF{$F \neq f_j$
\label{alg:skipLists:processProofComponent:checkNoComponents}}

\STATE Proof component is invalid

\ENDIF

\STATE $P \leftarrow \varnothing$

\FOR{$l = 0$ to $F$}\label{alg:skipLists:processProofComponent:loopStart}

\STATE $L \leftarrow h(j \| l \| d \| T_l)$ \COMMENT{Calculates
$L_j^l$. If the proof is correct, then $T_l$ must be $T^{j - 2^l}$ in
the original AASL.}
\label{alg:skipLists:processProofComponent:L}

\STATE $P \leftarrow P \| L$
\label{alg:skipLists:processProofComponent:concatenate}

\ENDFOR \label{alg:skipLists:processProofComponent:loopEnd}

\STATE $T \leftarrow h(P)$ \COMMENT{Should calculate $T^j$.}
\label{alg:skipLists:processProofComponent:T}

\STATE Return $T$

\end{algorithmic}
\end{algorithm}

\begin{algorithm}
\caption[Process a membership proof within an
AASL.]{\small\algsig{alg:skipLists:processWholeMembershipProof}{ProcessMembershipProof}
$(i, n, d, T, E) \Rightarrow \mathit{TRUE}/\mathit{FALSE}$.  Process the
membership proof $E$ of the membership claim $\langle i, n, d \rangle$
against authenticator $T$.}
\begin{algorithmic}[1]
\small

\STATE $\langle C_1, C_2, \ldots, C_S \rangle \leftarrow E$
\COMMENT{Parse $E$.}

\STATE $T_\mathit{cur} \leftarrow$
\algref{alg:skipLists:processProofComponent} $(i, C_1)$
\label{alg:skipLists:processWholeMembershipProof:firstStart}
\COMMENT{Should calculate $T^i$.}

\STATE $T_\mathit{prev} \leftarrow T_\mathit{cur}$

\STATE $l \leftarrow$ \algref{alg:skipLists:singleHopTraversalLevel}
$(i, n)$

\STATE $j \leftarrow i + 2^l$
\label{alg:skipLists:processWholeMembershipProof:firstEnd}
\label{alg:skipLists:processWholeMembershipProof:firstIncrement}

\STATE $c \leftarrow 2$ \COMMENT{Component counter.}

\WHILE{$j \leq n$}
\label{alg:skipLists:processWholeMembershipProof:loopStart}

\STATE $T_\mathit{cur} \leftarrow$
\algref{alg:skipLists:processProofComponent} $(j, C_c)$ \COMMENT{Should
return $T^j$.}
\label{alg:skipLists:processWholeMembershipProof:T}

\STATE $\langle d'; \langle T_0, T_1, \ldots, T_F \rangle \rangle
\leftarrow C_c$
\label{alg:skipLists:processWholeMembershipProof:continuityStart}
\label{alg:skipLists:processWholeMembershipProof:parseC}

\IF{$T_l \neq T_\mathit{prev}$}
\label{alg:skipLists:processWholeMembershipProof:continuity}

\STATE Proof is invalid \COMMENT{The values for the same authenticator
computed in the previous step and included in the current component
differ.}

\ENDIF \label{alg:skipLists:processWholeMembershipProof:continuityEnd}

\STATE $T_\mathit{prev} \leftarrow T_\mathit{cur}$
\label{alg:skipLists:processWholeMembershipProof:previousCurrent}

\STATE $l \leftarrow$ \algref{alg:skipLists:singleHopTraversalLevel}
$(j, n)$

\STATE $j \leftarrow j + 2^l$
\label{alg:skipLists:processWholeMembershipProof:increment}

\STATE $c \leftarrow c + 1$

\ENDWHILE \label{alg:skipLists:processWholeMembershipProof:loopEnd}

\IF{$S \neq c$}
\label{alg:skipLists:processWholeMembershipProof:checkNoComponents}

\STATE Proof is invalid \COMMENT{Wrong number of proof components.}

\ENDIF

\IF{$T_\mathit{cur} \neq T$}
\label{alg:skipLists:processWholeMembershipProof:checkAuthenticator}

\STATE Proof is invalid \COMMENT{The $T^n$ just computed from the proof
is different from the $T^n$ known.}

\ENDIF

\STATE $\langle d'; \langle \ldots \rangle \rangle \leftarrow C_1$
\COMMENT{Parse the datum in the first
component.}\label{alg:skipLists:processWholeMembershipProof:responseStart}

\IF{$d = d'$}
\label{alg:skipLists:processWholeMembershipProof:compareDatum}

\STATE Return TRUE

\ELSE

\STATE Return FALSE

\ENDIF \label{alg:skipLists:processWholeMembershipProof:responseEnd}

\end{algorithmic}
\end{algorithm}

Section~\ref{sec:skipLists:security} proves the security properties of
AASLs, as described informally above.  Namely, given an AASL digest
known to verifiers who follow
\algref{alg:skipLists:processWholeMembershipProof}, the maintainer can
only authenticate a single, unique membership claim per element to any
of those verifiers, and he can determine that digest only after he has
decided which claims he wishes to authenticate.

\subsection{\label{sec:skipLists:design:evolution}AASL Evolution}

AASLs are useful in distributing the contents of fixed-forever data
sequences, but can be invaluable in distributing the contents of data
sequences that grow over time.  In this section we address how AASLs can
be used when the data sequences on which they are based evolve over
time, especially when the verifier needs to access the sequence as it
changes.

As new elements are appended to a data sequence that a maintainer keeps
in an AASL, the AASL grows with new authenticators for the new elements.
Whenever it is necessary to commit to newer versions of the AASL, the
maintainer updates verifiers with the new AASL digest, i.e., the
currently last AASL authenticator.  In addition to the security
guarantees described in the previous section, verifiers of a dynamic
AASL must also be convinced that membership claims they verified in
previous versions of the AASL remain true in the new version.  Simply,
the AASL maintainer must be unable to ``rewrite history'' to which he has
committed in the past when he advances to a new version of the
structure.

The preservation of AASL history is supported by an \emph{advancement
proof}, which accompanies the new digest in an AASL version update.  An
advancement proof is very similar to a membership proof.  Intuitively,
an advancement proof authenticates a membership claim about an
authenticator, instead of a data value.  We call this an
\emph{advancement claim}; it has the form ``$T_\mathit{prev}$ is the
$i$-th authenticator of the AASL whose $n$-th authenticator is
$T_\mathit{new}$.''

\begin{figure}
\centerline{\includegraphics{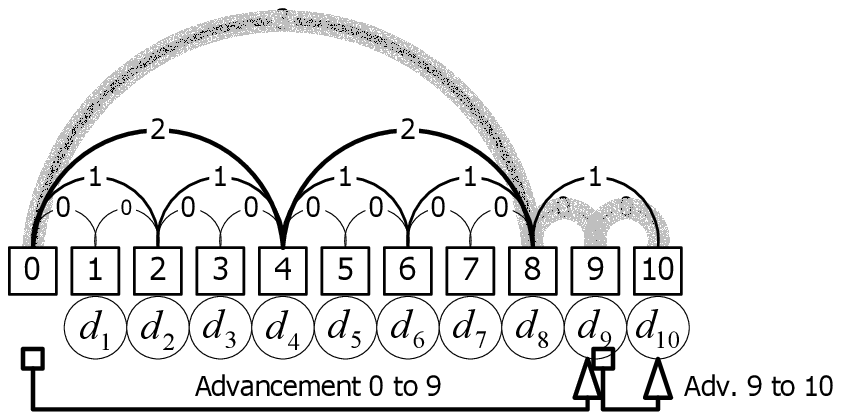}}
\caption[An example of advancement in a dynamic AASL]{An example of
advancement in a dynamic AASL. In version 1, the AASL has elements 1
through 9.  The corresponding advancement proof from the empty AASL to
version 1 is $A^{0,9}=\langle \langle d_8; \langle T^7, T^6, T^4, T^0
\rangle \rangle, \langle d_9; \langle T^8 \rangle \rangle \rangle$.
Then the maintainer adds element 10 and publishes version 2, with
advancement proof $A^{9,10}=\langle \langle d_{10}; \langle T^{9}, T^{8}
\rangle \rangle \rangle$.  The gray links delineate the traversal paths
that the two advancements take.}
\label{fig:skipLists:SingleAdvancement}
\end{figure}

Because advancement proofs are basically membership proofs, their
construction is almost identical to membership proof construction, and
their components have the same form.  Advancement proof $A^{i, n}$ from
the $i$-th to the $n$-th AASL authenticator is the same as membership
proof $E^{i, n, d}$ without the first proof component.  The first proof
component of a membership proof computes the authenticator of the source
element from the element datum and earlier AASL authenticators; this
step is unnecessary when the source element authenticator is already
known, as is the case with advancement.  In the appendix, we outline the
advancement proof construction algorithm
(Algorithm~\ref{alg:skipLists:constructWholeAdvancementProof}).
Figure~\ref{fig:skipLists:SingleAdvancement} illustrates an example of
advancement.

A verifier need remember three pieces of information for a given remote
dynamic AASL: the latest AASL size $n$, the latest digest $T$, and a
vector of earlier authenticators called a \emph{basis}.

The basis vector is used to check consistency among the values of
``reusable'' authenticators included in different advancement proofs for
the same AASL.  Reusable authenticators are those AASL authenticators
that may appear again in subsequent advancement proofs for the AASL.  In
the appendix, we illustrate an example of cheating that a malicious AASL
maintainer can perpetrate when he is free to use inconsistent values for
such reusable authenticators across advancements.  

The structure of a basis vector resembles the binary representation of
the AASL element index to which it corresponds.  Specifically, basis
$B^i$ for element $i$ is a vector of $l$ authenticators, where $l =
\lfloor \log_2 i \rfloor$ is the number of significant bits in the
binary representation of $i$. The vector contains a special ``empty''
value in those positions in which the binary representation of $i$
contains a 0; the rest of the basis vector's positions are occupied by
authenticator values.  These authenticator values correspond to the
authenticators of the elements encountered in the traversal of the AASL
from element 0 to element $i$.  A traversal from 0 to $i$ proceeds in
hops of decreasing length, starting with the largest power of 2 that is
less than or equal to the destination.  For example, for destination 9
(binary $1001$), the traversal from 0 first hops over $8 = 2^3$ elements
to element 8, and then over one last element ($2^0$) to element 9 (see
Figure~\ref{fig:skipLists:SingleAdvancement}).  In the associated basis
$B^9$, each non-zero ``bit'' position is annotated with the
authenticator of the element from which the corresponding traversal hop
launches, that is $B^9 = \langle T^0, \varnothing, \varnothing, T^8
\rangle$.  The basis $B^0$ for the $0$-th element (the initial value of
the AASL) contains no values.  Note that verifiers need not remember
bases for a static AASLs, since the concept of advancement is
meaningless in those.

Advancement proof verification occurs in two phases.  First, the
verifier checks whether the last digest he holds can appear in the AASL
of the new digest.  This check is almost identical to the verification
of membership proofs, as described in
Section~\ref{sec:skipLists:design:membership}, with the exception that
what is verified is the membership of an authenticator, not a datum, in
the AASL.

The second phase of checking an advancement proof deals with the basis.
For every component in the proof, the authenticators included therein
are checked against the values of any corresponding authenticators in
the basis.  If the component is consistent with remembered authenticator
values, the basis is updated with any reusable authenticators seen first
in the component.  In the end, the basis is updated to reflect the newly
acquired digest and advancement proof.
Algorithm~\ref{alg:skipLists:processAdvancementProofComponent} provides
the details, and is reminiscent of binary addition of positive integers.

\begin{algorithm}
\caption[Process an advancement component for a dynamic
AASL.]{\small\algsig{alg:skipLists:processAdvancementProofComponent}{ProcessAdvancementProofComponent}
$(j, T, B, C, l) \Rightarrow B'$.  Process an advancement component $C$
that takes a hop of level $l$ from the $j$-th digest $T$ with basis $B$.
Return the new basis.  }
\begin{algorithmic}[1]
\small

\STATE $\langle d; \langle T_0, T_1, \ldots, T_F \rangle \rangle
\leftarrow C$ \COMMENT{Parse $C$.}

\IF{$F \neq f_j$
\label{alg:skipLists:processAdvancementProofComponent:checkNoComponents}}

\STATE Proof component is invalid \COMMENT{The component contains the
wrong number of authenticators.}

\ENDIF

\STATE $\langle B_0, \ldots, B_b \rangle \leftarrow B$ \COMMENT{The
values in the basis vector.}

\IF{$B_l = \varnothing$}

\STATE $B_l \leftarrow T$
\label{alg:skipLists:processAdvancementProofComponent:or}

\STATE Return $\langle B_0, \ldots, B_b \rangle$

\ELSE

\STATE $c \leftarrow l$ \COMMENT{Current basis element.}

\WHILE{$B_c \neq \varnothing$}

\IF{$B_c \neq T_{c + 1}$}
\label{alg:skipLists:processAdvancementProofComponent:consistency}

\STATE Advancement is invalid. \COMMENT{The maintainer is now sending a
different value ($T_{c + 1}$) for an authenticator whose value he
reported as $B_c$ before.}

\ENDIF

\STATE $\mathit{carry} \leftarrow B_c$

\STATE $B_c \leftarrow \varnothing$
\label{alg:skipLists:processAdvancementProofComponent:zero}

\STATE $c \leftarrow c + 1$

\ENDWHILE

\STATE $B_c \leftarrow \mathit{carry}$

\STATE Return $\langle B_0, \ldots, B_{\max\{b,c\}} \rangle$
\COMMENT{The vector may have grown by one non-empty element.}

\ENDIF

\end{algorithmic}
\end{algorithm}

\algref{alg:skipLists:processAdvancementProofComponent} is invoked once
for every component in the advancement proof, after that component has
been processed as a membership proof component.
Algorithm~\ref{alg:skipLists:processWholeAdvancementProof} describes how
the whole advancement proof verification proceeds.

\begin{algorithm}
\caption[Process an advancement proof within an
AASL.]{\small\algsig{alg:skipLists:processWholeAdvancementProof}{ProcessAdvancementProof}
$(i, n, T_\mathit{prev}, B_\mathit{prev}, T_\mathit{new}, A) \Rightarrow
B_\mathit{new}$.  Process the advancement proof $A$ that establishes
$T_\mathit{new}$ as the $n$-th authenticator, starting with the $i$-th
authenticator $T_\mathit{prev}$ and basis $B_\mathit{prev}$.  The
process returns the new basis $B_\mathit{new}$, if successful.}
\begin{algorithmic}[1]
\small

\STATE $\langle C_2, \ldots, C_S \rangle \leftarrow A$ \COMMENT{Parse
$A$.  The numbering starts with 2, to be consistent with the numbering
in \algref{alg:skipLists:processWholeMembershipProof}.}

\STATE $c \leftarrow 2$ \COMMENT{Component counter.}

\STATE $j \leftarrow i$
\label{alg:skipLists:processWholeAdvancementProof:firstEnd}

\WHILE{$j < n$}
\label{alg:skipLists:processWholeAdvancementProof:loopStart}

\STATE $l \leftarrow$ \algref{alg:skipLists:singleHopTraversalLevel}
$(j, n)$ \label{alg:skipLists:processWholeAdvancementProof:l}

\STATE $B_\mathit{new} \leftarrow$
\algref{alg:skipLists:processAdvancementProofComponent} $(j,
T_\mathit{prev}, B_\mathit{prev}, C_c, l)$ \COMMENT{This returns $B^{j +
2^l}$.}\label{alg:skipLists:processWholeAdvancementProof:B}

\STATE $j \leftarrow j + 2^l$ \COMMENT{Next element in traversal.}
\label{alg:skipLists:processWholeAdvancementProof:increment}

\STATE $T_\mathit{cur} \leftarrow$
\algref{alg:skipLists:processProofComponent} $(j, C_c)$ \COMMENT{Should
be $T^j$.}
\label{alg:skipLists:processWholeAdvancementProof:T}

\STATE $\langle d; \langle T_0, T_1, \ldots, T_F \rangle \rangle
\leftarrow C_c$ \COMMENT{Parse $C_c$.}
\label{alg:skipLists:processWholeAdvancementProof:continuityStart}
\label{alg:skipLists:processWholeAdvancementProof:parseC}

\IF[$T_\mathit{prev}$ should be $T^{j - 2^l}$.]{$T_l \neq
T_\mathit{prev}$}
\label{alg:skipLists:processWholeAdvancementProof:continuity}

\STATE Proof is invalid \COMMENT{The value of $T^{j - 2^l}$ computed in
the previous step is not the same as the value for $T^{j - 2^l}$ in the
current proof component.}

\ENDIF \label{alg:skipLists:processWholeAdvancementProof:continuityEnd}

\STATE $T_\mathit{prev} \leftarrow T_\mathit{cur}$
\label{alg:skipLists:processWholeAdvancementProof:previousCurrent}

\STATE $B_\mathit{prev} \leftarrow B_\mathit{new}$

\STATE $c \leftarrow c + 1$

\ENDWHILE \label{alg:skipLists:processWholeAdvancementProof:loopEnd}

\IF{$S \neq c$}
\label{alg:skipLists:processWholeAdvancementProof:checkNoComponents}

\STATE Proof is invalid \COMMENT{Wrong number of proof components.}

\ENDIF

\IF{$T_\mathit{cur} \neq T_\mathit{new}$}
\label{alg:skipLists:processWholeAdvancementProof:checkAuthenticator}

\STATE Proof is invalid \COMMENT{The $T^n$ claimed by the advancement is
different from the one computed by processing the advancement proof.}

\ENDIF

\STATE Return $B_\mathit{new}$

\end{algorithmic}
\end{algorithm}

A powerful use of AASLs is to determine the possible relative orders of
insertion of different data in the maintainer's tamper-evident data
sequence.  For example, let Molly by an AASL maintainer who claims that
she did not learn value $a$ until after she had committed to value $b$.
If verifier Van holds valid proofs of the membership claims $\langle i,
j, a \rangle$ and $\langle k, n, b \rangle$ in Molly's AASL, where $i <
j < k < n$, then he can convince anyone who agrees on Molly's $j$-th and
$n$-th AASL authenticators that she is lying; Molly must have known
value $a$ before her commitment to the $j$-th authenticator, and
therefore before her commitment to $b$.  Such temporal orderings can
apply also to the data themselves, when those data contain a ``freshness
marker'', as is the case, for example, with signed statements containing
a nonce.  We detail how temporal ordering in a distributed log can be
preserved in the Timeweave project~\cite{Maniatis2002b}.

In the next section, we prove the security properties of static AASLs,
described in Section~\ref{sec:skipLists:design:membership}, and of
dynamic AASLs, described in this section.

\section{\label{sec:skipLists:security}Security Analysis}

In this section, we substantiate the security guarantees that AASLs
offer to their users.  Our goal is to secure the ``commitment metaphor''
of AASLs for verifiers who follow the membership and advancement proof
verification procedures described in the previous section.  Informally,
this means that, first, diligent verifiers accept only a single, unique
membership claim for every position in the data sequence on which an
AASL is built; second, the data structure maintainer must decide which
membership claims he can prove before he commits to the AASL by giving a
digest to potential verifiers.

There are two distinct ``roles'' that a malicious adversary can take,
with regards to an AASL.  On one hand, the adversary may be an
eavesdropper, who wishes to prove to a verifier a false membership claim
of his choosing, for an AASL that he does not maintain.  On the other
hand, the adversary may be the AASL maintainer, who wishes either to
defer choosing to which membership claim to commit until after he has
apparently committed; or to prove conflicting membership claims to
different verifiers (a membership claim $\langle i, n, d \rangle$
conflicts with membership claim $\langle i, n', d' \rangle$ if $d \neq
d'$).  A malicious AASL maintainer is a more powerful adversary, because
he can use arbitrary means to produce a digest before he has to relay it
to potential verifiers.  In what follows, we prove that AASLs are
resistant to such attacks.

First, we show that an adversary is unable to construct convincing
membership proofs (that he has not already seen) from a random AASL
digest.  This prevents a malicious eavesdropper from proving false
membership claims.  This also prevents a malicious AASL maintainer from
committing to bogus digests and only deciding later what to prove to its
unsuspecting verifiers.  This property is similar to the pre-image
resistance property of one-way functions.

\begin{theorem}[AASL Membership Proof Pre-image Resistance]
\label{thm:skipLists:proofPreImageResistance}
Consider randomly chosen $T$ from the set of values of the hash function
$h$.  A computationally bound adversary cannot construct efficiently an
AASL membership proof $E^{i,n,d}$ of any datum $d$ in position $i$ of an
$n$-element AASL, for any $i$ and $n$ ($0 < i \leq n$).
\end{theorem}

This result follows directly from the pre-image resistance of the hash
function $h$.

Suppose the adversary can pick $d$, $i$ and $n$ and construct a
membership proof $E^{i,n,d}$ of $d$ in position $i$, where $T$ is the
given $n$-th authenticator, so that a verifier in possession of $T$ and
following Algorithms~\ref{alg:skipLists:processProofComponent} and
\ref{alg:skipLists:processWholeMembershipProof} accepts the proof.

Given $E^{i,n,d}$, $i$, $d$, $n$ and $T$,
Algorithm~\ref{alg:skipLists:processWholeMembershipProof} executed by
the verifier must fail to match the condition of
Line~\ref{alg:skipLists:processWholeMembershipProof:checkAuthenticator}.
This means that in the last iteration of
Line~\ref{alg:skipLists:processWholeMembershipProof:T}, $T_\mathit{cur}$
returned from Algorithm~\ref{alg:skipLists:processProofComponent} must
be the random $T$ given to the adversary in the challenge.  However,
this means that, in Line~\ref{alg:skipLists:processProofComponent:T} of
Algorithm~\ref{alg:skipLists:processProofComponent}, the adversary must
be able to find a pre-image of the pre-image resistant hash function $h$
for random image $T$.  As a result, the hypothesis is false, and the
adversary cannot produce a pre-image proof.\qed

Theorem~\ref{thm:skipLists:proofPreImageResistance} only deals with
cheap, unsophisticated malice.  We proceed by addressing more
sophisticated attacks that rely on the manipulation of corrupt AASLs by
their maintainer or on the manipulation of observed proofs by an
eavesdropper.  There are three types of such attacks.  First, the
adversary can modify correct proofs to make them prove a false
membership claim.  Second, the maintainer can produce an AASL digest
against which he can prove conflicting membership claims.  Third, the
maintainer can produce AASL digests and advancement proofs so as to
prove conflicting membership claims against different versions of the
AASL.  We call the first two attacks second pre-image and collision,
respectively.  We call the third attack \emph{evolutionary collision},
because it relies on subverting AASL evolution across versions.

In the next theorem, we prove that AASLs are resistant to the second
type of attack, collision attacks
(Theorem~\ref{thm:skipLists:proofCollisionResistance}).  AASLs are also
resistant to the first type of attack, second pre-image, but the proof
is a direct corollary of collision resistance, so we defer to the
Appendix for it
(Theorem~\ref{thm:skipLists:proofSecondPreImageResistance}).

\begin{theorem}[AASL Membership Proof Collision Resistance]
\label{thm:skipLists:proofCollisionResistance}

A computationally bound adversary cannot construct two membership proofs
$E$ and $E'$ verifiable against the same authenticator $T$ that
authenticate different data values in the same sequence position.

\end{theorem}

Suppose that an adversary can, in fact, construct an efficient proof
collision with proofs $E$ and $E'$ against common authenticator $T$.
Let the two membership claims be $t = \langle i, n, d \rangle$ and $t' =
\langle i, n', d' \rangle$, respectively ($t \neq t'$).  We trace
\algref{alg:skipLists:processWholeMembershipProof} backwards for both
proofs $E$ and $E'$ in parallel, and reach a violation of the one-way
properties of the hash function $h$.

Since both membership proofs can be verified against the same
authenticator $T$ (which corresponds to a purported AASL's $n$-th
element in the case of $E$ and a different purported AASL's $n'$-th
element in the case of $E'$), in the last iteration of
Line~\ref{alg:skipLists:processWholeMembershipProof:T} of
Algorithm~\ref{alg:skipLists:processWholeMembershipProof}, the
invocation of Algorithm~\ref{alg:skipLists:processProofComponent} must
yield the same result $T$.  In this last iteration, local variable $j$,
the current element of the purported AASL, is equal to $n$ and $n'$,
respectively.

However, this means that the adversary must be able to cause the
verifier to invoke \algref{alg:skipLists:processProofComponent} with
input $(n, C)$ and $(n', C')$ but receive the same result $T$ for both
invocations.  This is equivalent to passing to
Equations~\ref{eqn:skipLists:L} and \ref{eqn:skipLists:T} different
$i$'s and $T$'s but calculating the same $T^i$.  Intuitively, since the
two equations use a one-way hash function, this should be impossible,
i.e., \algref{alg:skipLists:processProofComponent} should only return
the same result when invoked with identical inputs (we prove this
rigorously in the Appendix, in
Lemma~\ref{lem:skipLists:differentProofComponentsSameT}).  Therefore, in
the last iteration \algref{alg:skipLists:processProofComponent} can only
be invoked with $(n, C)$ and $(n', C')$ if $n = n'$.  This restricts our
assumed proof collision to support membership claims that only differ in
the data values $d$ and $d'$.

Since both proofs authenticate position $i$ in an $n$-length AASL,
Line~\ref{alg:skipLists:processWholeMembershipProof:checkNoComponents}
of \algref{alg:skipLists:processWholeMembershipProof} imposes that the
proof lengths must be equal to the same $S$.  We prove inductively on
the number of components in the two proofs that the two proofs must be
identical.  Induction follows the iterations of the loop in
\algref{alg:skipLists:processWholeMembershipProof},
Lines~\ref{alg:skipLists:processWholeMembershipProof:loopStart} --
\ref{alg:skipLists:processWholeMembershipProof:loopEnd}, from last
iteration to first.

The base case for the last components $C_S$ and $C'_S$, respectively,
follows directly from the collision resistance claim of
\algref{alg:skipLists:processProofComponent}
(Lemma~\ref{lem:skipLists:differentProofComponentsSameT} in the
appendix) and from the supposition that both proofs are verifiable
against the same authenticator $T$.

To establish the inductive step, consider the $c$-th proof components
$C_c$ and $C'_c$ of the two membership proofs and assume they are equal.
In the associated loop iteration in
\algref{alg:skipLists:processWholeMembershipProof},
Line~\ref{alg:skipLists:processWholeMembershipProof:parseC} extracts the
individual $F$ hash values of the $c$-th proof component; these are
pairwise equal across the two respective proof components, since the
components themselves are equal.  The $l$-th of these hash values must
be equal to the value of the respective $T_\mathit{prev}$, in
Line~\ref{alg:skipLists:processWholeMembershipProof:continuity}.  Since
the $l$-th hash values are equal across proofs, the values of
$T_\mathit{prev}$ are the same in the invocations of
\algref{alg:skipLists:processWholeMembershipProof} for the two
membership proofs.  But, in the previous loop iteration, in
Line~\ref{alg:skipLists:processWholeMembershipProof:previousCurrent},
$T_\mathit{prev}$ had been assigned the value of the respective
$T_\mathit{cur}$, computed using
\algref{alg:skipLists:processProofComponent} in
Line~\ref{alg:skipLists:processWholeMembershipProof:T}.  Because of the
collision resistance of \algref{alg:skipLists:processProofComponent},
this means that the inputs to the two respective invocations of
\algref{alg:skipLists:processProofComponent} must also be identical in
that loop iteration, which means that the $(c-1)$-st element components
$C_{c-1}$ and $C'_{c-1}$, respectively, are also identical.  This proves
the inductive step.

The induction applies to all but the first proof components in the two
proofs, which are processed outside the loop of
\algref{alg:skipLists:processWholeMembershipProof}, in
Lines~\ref{alg:skipLists:processWholeMembershipProof:firstStart} --
\ref{alg:skipLists:processWholeMembershipProof:firstEnd}.  The same
argument as the inductive step above can also be applied here: the
$T_\mathit{cur}$ returned by the respective invocations of
\algref{alg:skipLists:processProofComponent} on the respective first
proof components is the same $T_\mathit{prev}$ that ends up matching the
identical $l$-th hash values of the respective, equal second proof
components in
Line~\ref{alg:skipLists:processWholeMembershipProof:continuity} of the
first loop iteration.  Consequently, the respective first proof
components must also be equal.

We have shown that two proofs $E$ and $E'$ authenticating the same
element position $i$ against the same authenticator $T$ must be
identical.  But in \algref{alg:skipLists:processWholeMembershipProof},
Line~\ref{alg:skipLists:processWholeMembershipProof:compareDatum}, the
datum in the first component of a proof must match the one whose
membership is verified.  This contradicts the collision hypothesis,
because the condition in
Line~\ref{alg:skipLists:processWholeMembershipProof:compareDatum} only
succeeds if the algorithm is invoked with the data value that occupies
the first proof component of the two proofs.  $d$ and $d'$ cannot be
different.  \qed

Finally, we prove that AASLs are resistant to the third type of
malicious manipulation attack, evolutionary collision, in
Theorem~\ref{thm:skipLists:evolutionaryCollisionResistance}.  AASLs have
the property of evolutionary collision-resistance if it is impossible
for a computationally constrained adversary to produce advancements and
membership proofs that authenticate two different data elements $d$ and
$d' \neq d$ for the same position $i$, in \emph{any} version of the same
AASL.

The definition is fairly broad in scope: it covers unrelated, mutually
unknown verifiers $\mathcal{A}$ and $\mathcal{B}$, who, through
different sequences of advancements, arrive at the same digest $T$ for
position $n$ of an AASL at different times; a malicious prover must be
unable to convince $\mathcal{A}$ that $d$ is at position $i$ \emph{and}
convince $\mathcal{B}$ that $d' \neq d$ is at position $i$, even in
different versions of the AASL in its separate evolution paths towards
length $n$ and digest $T$.

Two advancements $A^{i,j}$ and $A^{k,l}$ are \emph{connected} if the
source element of the latter advancement is the destination element of
the former, that is $j = k$.  In what follows, we refer to a sequence of
connected advancements as an \emph{advancement sequence}, and the
sequence of element positions traversed by that advancement sequence as
an \emph{advancement path}.  In a similar manner, we define the sequence
of element positions traversed by a membership proof as a
\emph{membership proof path}.

Our proof strategy for evolutionary collision resistance is to show that
if two diligent verifiers have both accepted the same authenticator for
the same AASL element, they must arrive at the same value for the
authenticators of some other strategic AASL elements.  Namely, we show
that the two verifiers must ``agree'' on the authenticators they compute
during the processing of the membership proofs with which the adversary
seeks to fool them.  From
Theorem~\ref{thm:skipLists:proofCollisionResistance}, if two verifiers
agree on the authenticators computed during membership proof
verification, they cannot be verifying the truth of conflicting
membership claims.

To reduce authenticator agreement during the verification of independent
advancement paths to authenticator agreement during the verification of
independent membership proofs, we use two ``authenticator agreement
claims,'' which we describe here informally, but prove rigorously in the
Appendix.

First, if a membership proof verification and an advancement proof
verification agree on the value of a particular AASL authenticator, then
they must also agree on the authenticator values of all earlier AASL
elements that the two paths---the advancement and the membership proof
paths---have in common (see
Lemma~\ref{lem:skipLists:authenticatorAgreementProofAdvancement} in the
appendix).

Second, if two runs of the advancement verification algorithm, applied
to two different advancement sequences, agree on the value of a
particular AASL authenticator, then they must also agree on the
authenticator values of all earlier AASL elements that the two
advancement paths have in common (see
Lemma~\ref{lem:skipLists:commonAdvancementPoints} in the appendix).

Equipped with these two claims, we now tackle evolutionary collision
resistance.

\begin{theorem}[Evolutionary collision resistance of AASL membership
proofs.]
\label{thm:skipLists:evolutionaryCollisionResistance}
Consider two independent verifiers, $\mathcal{A}$ and $\mathcal{B}$ and
a computationally constrained adversary who conveys to them
independently two advancement sequences.  It is impossible for the
adversary to produce advancement sequences and membership proofs in such
a way that, first, the two verifiers, processing their respective
advancement sequences, advance to element position $n$ with the same
digest $T$; and, second, the two verifiers, processing separate
membership proofs, authenticate, at any time, two conflicting membership
claims.
\end{theorem}

It is already known, from
Theorem~\ref{thm:skipLists:proofCollisionResistance}, that conflicting
membership claims cannot be authenticated against the \emph{same}
authenticator.  Here we address the case where the two aspiring proofs
authenticate different data values for the same AASL position against
the authenticators of different versions of that AASL held by the two
verifiers.

Let $i$ be the element position for whose data element the adversary
wishes to fool two verifiers, $\mathcal{A}$ and $\mathcal{B}$, and let
$j$ and $k$ be the element positions against whose authenticators he
wishes to produce the offending proofs for the verifiers; specifically,
the adversary wishes to authenticate the membership claims $\langle i,
j, d \rangle$ to $\mathcal{A}$ and $\langle i, k, d' \rangle$ to
$\mathcal{B}$.  Without loss of generality, we assume $j < k$, so $0 < i
\leq j < k \leq n$.

\begin{figure}
\centerline{\includegraphics{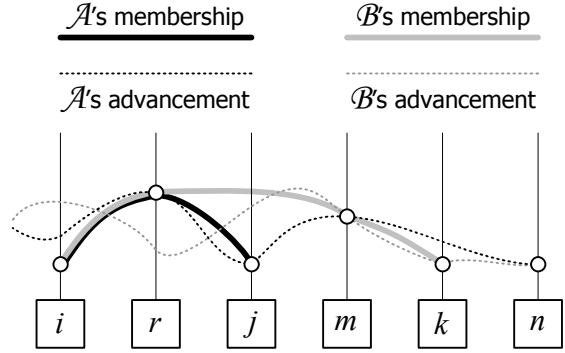}}
\caption[Common elements in independent verifiers' view of the same
AASL.]{Illustration of the proof of
Theorem~\ref{thm:skipLists:evolutionaryCollisionResistance}. Verifier
$\mathcal{A}$ advances to the $n$-th digest of the AASL via element $j$.
When $\mathcal{A}$ held the $j$-th digest for the AASL, he had
successfully authenticated a datum for the $i$-th position.  Verifier
$\mathcal{B}$ advances to the $n$-th digest of the AASL via element $k$.
When $\mathcal{B}$ held the $k$-th digest for the AASL, he had
successfully authenticated a datum for the same $i$-th position as
$\mathcal{A}$ did.}
\label{fig:skipLists:CommonElements}
\end{figure}

Consider the abstract illustration of this setup in
Figure~\ref{fig:skipLists:CommonElements}.  $\mathcal{A}$'s advancement
path, the dark dashed line, does not necessarily go through element $i$,
but it certainly touches element $j$ (since the adversary's membership
proof is authenticated to $\mathcal{A}$ against the $j$-th
authenticator) and element $n$ (since the two verifiers agree on the
value of the $n$-th authenticator).  Similarly, $\mathcal{B}$'s
advancement path, the lighter dashed line, does not necessarily go
through element $i$, but certainly touches elements $k$ and $n$.
$\mathcal{A}$'s membership proof path (the thick dark line) starts from
$i$ and ends at $j$, and $\mathcal{B}$'s membership proof path (the
lighter dark line) starts from $i$ and ends at $k$.

There is an element in $[j, k]$, element $m$, that is common among
$\mathcal{A}$'s advancement path, $\mathcal{B}$'s advancement path, and
$\mathcal{B}$'s membership proof path. This results from the fact that
$\mathcal{B}$'s membership proof and advancement paths both start before
element $j$ and touch element $k$, and $\mathcal{A}$'s advancement path
touches element $j$ and continues past element $k$.  An intuitive reason
for this is that $\mathcal{A}$'s advancement path can skip element $k$
only by ``jumping'' over it on a high-level linked list.  Then,
$\mathcal{B}$'s membership proof and advancement paths must touch the
jumping-off point of $\mathcal{A}$'s path, on their way to ``lower''
$k$.  We prove this claim rigorously in
Lemma~\ref{lem:skipLists:parallelPaths}, in the Appendix.

The two advancements agree on the value of $T^n$ after processing the
respective advancement sequences in
\algref{alg:skipLists:processWholeAdvancementProof}, as per the theorem
assumption.  Because of the second authenticator agreement claim
described above, this means that the two advancement algorithms also
agree with each other on the value of $T^m$ after processing the
corresponding part of their respective advancement sequences that brings
them both to element $m$.

Because $\mathcal{B}$'s membership proof verification, to succeed, must
agree on the value of $T^k$ with the advancement verification algorithm,
and because of the first authenticator agreement claim above, the
membership proof verification algorithm on $\mathcal{B}$ also agrees on
the value of $T^m$ with $\mathcal{B}$'s advancement verification
algorithm after they both reach element $m$.  Therefore, $\mathcal{B}$'s
membership proof verification algorithm and both advancement
verification algorithms agree on the value of $T^m$ after reaching
element $m$.

As above, there is an element in $[i, j]$, element $r$, that is common
among $\mathcal{A}$'s advancement path, and $\mathcal{A}$ and
$\mathcal{B}$'s membership proof paths.  This is because both membership
proof paths start at $i$ and go to or past $j$, and $\mathcal{A}$'s
advancement path starts before $i$ and touches element $j$ (since
$\mathcal{A}$'s membership proof must be verifiable against the digest
for element $j$, as per the theorem assumptions).

Because $\mathcal{A}$'s membership and advancement proof verification
algorithms must agree on the value of $T^j$ for the membership proof to
be accepted, and from the first authenticator agreement claim once more,
the membership proof verification algorithm on $\mathcal{A}$ also agrees
on the value of $T^r$ with $\mathcal{A}$'s advancement algorithm after
reaching element $r$.  From the same claim, since $\mathcal{B}$'s
membership and $\mathcal{A}$'s advancement proof verification algorithms
agree on the value of $T^m$ after reaching element $m$, they must also
agree on the value of $T^r$ after they reach element $r$.  As a result,
the two membership proof verification algorithms reach element $r$ with
the same value for $T^r$.

However, this contradicts
Theorem~\ref{thm:skipLists:proofCollisionResistance}.  If the adversary
could manage to create two membership proofs starting with different
data values on element $i$ and computing the same authenticator for
element $r$, then he would be able to produce same-version collisions,
as well, which Theorem~\ref{thm:skipLists:proofCollisionResistance}
precludes.  Therefore, the two data elements $d$ and $d'$ cannot be
different.\qed

\section{Conclusions}

In this work we describe, design and analyze the security of a
tamper-evident, append-only data structure for maintaining secure data
sequences in a loosely coupled distributed system, where individual
system components may be mutually distrustful.  The resulting data
structure, called Authenticated Append-Only Skip List, allows its
maintainers to produce one-way digests of the entire data sequence,
which they can publish to others as a commitment on the contents and
order of the sequence.  The maintainer can produce efficiently succinct
proofs that authenticate a particular datum in a particular position of
the data sequence against a published digest.

AASLs are secure against tampering even by malicious structure
maintainers.  First, we have shown that a maintainer cannot ``invent''
and authenticate data elements for the AASL after he has committed to
the structure.  Second, he cannot equivocate by being able to prove
conflicting facts about a particular position of the data sequence.
This is the case, even when the data sequence grows with time and its
maintainer publishes successive commitments at times of his own
choosing.

We have implemented and extensively measured the performance and storage
requirements of AASLs (we present a discussion of practical
implementation considerations in the Appendix).  We have used AASLs
extensively in Timeweave~\cite{Maniatis2002b}, a system for preserving
historic integrity in trust-free peer-to-peer systems.

{\small
\bibliographystyle{acm}
\bibliography{IdentiScape}

\begin{thebibliography}{1}

\bibitem{Anagnostopoulos2001}
{\sc Anagnostopoulos, A., Goodrich, M.~T., and Tamassia, R.}
\newblock {Persistent Authenticated Dictionaries and Their Applications}.
\newblock In {\em Proceedings of the Information Security Conference (ISC
  2001)\/} (Malaga, Spain, Oct. 2001), vol.~2200 of {\em Lecture Notes in
  Computer Science}, Springer, pp.~379--393.

\bibitem{Buldas1998}
{\sc Buldas, A., Laud, P., Lipmaa, H., and Villemson, J.}
\newblock {Time-stamping with Binary Linking Schemes}.
\newblock In {\em Advances on Cryptology (CRYPTO 1998)\/} (Santa Barbara, USA,
  Aug. 1998), H.~Krawczyk, Ed., vol.~1462 of {\em Lecture Notes in Computer
  Science}, Springer, pp.~486--501.

\bibitem{Goodrich2001}
{\sc Goodrich, M.~T., Tamassia, R., and Schwerin, A.}
\newblock {Implementation of an Authenticated Dictionary with Skip Lists and
  Commutative Hashing}.
\newblock In {\em 2001 DARPA Information Survivability Conference and
  Exposition (DISCEX 2001)\/} (Anaheim, CA, USA, June 2001).

\bibitem{Haber1991}
{\sc Haber, S., and Stornetta, W.~S.}
\newblock {How to Time-stamp a Digital Document}.
\newblock {\em Journal of Cryptology: the Journal of the International
  Association for Cryptologic Research 3}, 2 (1991), 99--111.

\bibitem{Maniatis2002b}
{\sc Maniatis, P., and Baker, M.}
\newblock {Secure History Preservation Through Timeline Entanglement}.
\newblock In {\em Proceedings of the 11th USENIX Security Symposium\/} (San
  Francisco, {CA}, {USA}, Aug. 2002), pp.~297--312.

\bibitem{SHA1}
{\sc National Institute of Standards and Technology ({NIST})}.
\newblock {\em {Federal Information Processing Standard Publication 180-1:
  Secure Hash Standard}}.
\newblock Washington, {D.C.}, {USA}, Apr. 1995.

\bibitem{Pugh1990}
{\sc Pugh, W.}
\newblock {Skip Lists: a Probabilistic Alternative to Balanced Trees}.
\newblock {\em Communications of the ACM 33}, 6 (June 1990), 668--676.

\bibitem{Schneier1998}
{\sc Schneier, B., and Kelsey, J.}
\newblock {Cryptographic Support for Secure Logs on Untrusted Machines}.
\newblock In {\em Proceedings of the 7th USENIX Security Symposium\/} (San
  Antonio, {TX}, {USA}, Jan. 1998), pp.~53--62.

\bibitem{Spreitzer1997}
{\sc Spreitzer, M.~J., Theimer, M.~M., Petersen, K., Demers, A.~J., and Terry,
  D.~B.}
\newblock {Dealing with Server Corruption in Weakly Consistent, Replicated Data
  Systems}.
\newblock In {\em Proceedings of the Third Annual {ACM/IEEE} International
  Conference on {M}obile Computing and Networking\/} (Budapest, Hungary, Sept.
  1997), ACM/IEEE, pp.~234--240.

\end{thebibliography}
}

\clearpage

\appendix
\section{The Need For Bases}

\begin{figure}
\centerline{\includegraphics{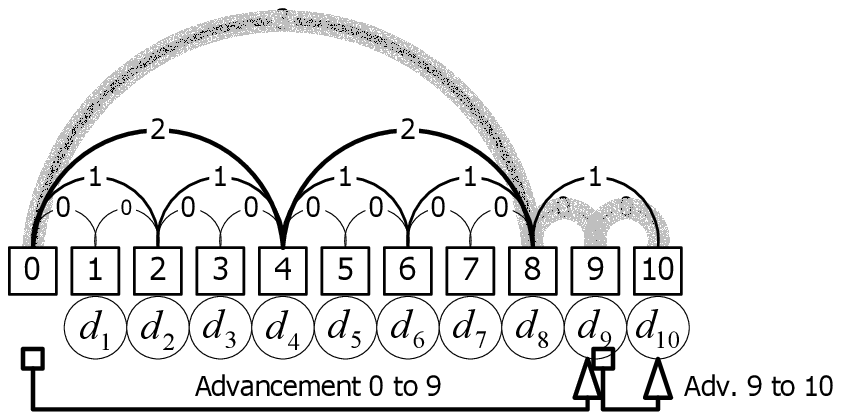}}
\caption[An example of advancement in a dynamic AASL]{A repeat of
Figure~\ref{fig:skipLists:SingleAdvancement}. An example of advancement
in a dynamic AASL. In version 1, the AASL has elements 1 through 9.  The
corresponding advancement is $A^{0,9}=\langle \langle d_8; \langle T^7,
T^6, T^4, T^0 \rangle \rangle, \langle d_9; \langle T^8 \rangle \rangle
\rangle$.  Then the maintainer adds element 10 and publishes version 2,
with advancement $A^{9,10}=\langle \langle d_{10}; \langle T^{9}, T^{8}
\rangle \rangle \rangle$.  The gray links delineate the traversal paths
that the two advancements take.}
\label{fig:skipLists:SingleAdvancement2}
\end{figure}

We give here a simple example of how ``forgetting'' the values of
reusable authenticators can allow a malicious maintainer to authenticate
conflicting membership claims across AASL versions.  Consider the
authenticator for element 8, in
Figure~\ref{fig:skipLists:SingleAdvancement2}; it is used in all
membership proofs verifiable against the digest of version 1 ending with
element 9, since the authenticator for element 9 depends on a single
partial authenticator, that for element 8.  However, the authenticator
for element 8 is also used in all membership proofs verifiable against
the digest of version 2 ending with element 10, because the
authenticator for element 10 also depends on the authenticator for
element 8 for one of its partial authenticators.

A malicious maintainer can construct two authenticators $T^8$ and
${T^8}'$ for element 8 to accommodate two different elements $d_8$ and
$d_8'$, respectively, using Equations~\ref{eqn:skipLists:L} and
\ref{eqn:skipLists:T}, as follows:
\begin{eqnarray}
T^8	& =	&	h(h(8 \| 0 \| d_8 \| T^7) \ \| \ 
			h(8 \| 1 \| d_8 \| T^6) \ \| \nonumber \\
	&	&	h(8 \| 2 \| d_8 \| T^4) \ \| \ 
			h(8 \| 3 \| d_8 \| T^0)) \nonumber \\
{T^8}'	& =	&	h(h(8 \| 0 \| d_8' \| T^7) \ \| \ 
			h(8 \| 1 \| d_8' \| T^6) \ \| \nonumber \\ 
	&	&	h(8 \| 2 \| d_8' \| T^4) \ \| \ 
			h(8 \| 1 \| d_8' \| T^0)) \nonumber
\end{eqnarray}
He can then construct a single authenticator $T^9$ for element 9 based
on ${T^8}'$:
\[ T^9 = h(h(9 \| 0 \| d_9 \| {T^8}')) \]
and use it to commit to version 1, which ends at element 9, with this
$T^9$ and the first advancement proof $A^{0,9}$:
\[ A^{0,9} = \langle \langle d_8'; \langle T^7, T^6, T^4, T^0 \rangle \rangle, \langle
d_9; \langle {T^8}' \rangle \rangle \rangle \] The digest $T^9$ for version 1
authenticates $d_8'$ in position 8 with the following membership proof:
\[ E^{8, 9, d_8'} = \langle \langle d_8'; \langle T^7, T^6, T^4, T^0 \rangle \rangle, \langle d_9;
\langle {T^8}' \rangle \rangle \rangle \]

Now the malicious maintainer can construct a corrupt authenticator
$T^{10}$ for the 10-th element, by mixing $T^8$ from the AASL that
contains $d_8$ in position 8, and ${T^9}$, from the AASL that contains
$d_8'$ in position 8:
\[ T^{10} = h(h(10 \| 0 \| d_{10} \| T^9) \| h(10 \| 1 \| d_{10} \|
T^8)) \] and publish it as the digest for version 2, with the
corresponding advancement proof
\[ A^{9, 10} = \langle \langle d_{10}; \langle
T^9, T^8 \rangle \rangle \rangle \] In conflict to version 1, version 2
authenticates element $d_8$ in position 8, with the following membership
proof:
\[ E^{8, 10, d_8} = \langle \langle d_8; \langle T^7, T^6, T^4, T^0 \rangle \rangle,
\langle d_{10}; \langle T^9, T^8 \rangle \rangle \rangle \]

The problem lies in the verifier's forgetting that the value for the
purported authenticator of element 8 was ${T^8}'$ in the first
advancement $A^{0,9}$ to version 1, whereas the same authenticator has
the value $T^8 \neq {T^8}'$ in the second advancement $A^{9,10}$ from
version 1 to version 2.  To avoid this problem, verifiers keep track of
reusable authenticators, such as $T^8$ in the example above.  With every
advancement received, a verifier checks that any reused authenticators
in the advancement agree with those known so far in the basis for the
same AASL; then, the verifier updates that basis with any new reusable
authenticators included in the newly received advancement.

\section{Algorithms}

Here we describe the proof construction algorithms in detail.  They do
not participate in any of the security proofs, since the security
guarantees offered by AASLs have to do with what claims diligent
verifiers accept.

First, Algorithm~\ref{alg:skipLists:singleElementComponent}, describes
how individual proof components are constructed.  These proof components
participate in both membership and advancement proofs.

\begin{algorithm}
\caption[Construction of a proof component for an
AASL.]{\small\algsig{alg:skipLists:singleElementComponent}{SingleProofComponent}
$(j) \Rightarrow C$.  Return a proof component $C$ for the AASL element
in position $j$.}
\begin{algorithmic}[1]
\small

\STATE $T_\mathit{vec} \leftarrow \varnothing$ \COMMENT{Authenticators.}

\FOR{$l = 0$ to $f_j$}

\STATE $T_\mathit{vec} \leftarrow T_\mathit{vec} \| T^{j - 2^l}$

\ENDFOR

\STATE $C \leftarrow \langle d_j; T_\mathit{vec} \rangle$

\STATE Return $C$

\end{algorithmic}
\end{algorithm}

Then, we proceed by describing how a whole membership proof
(Algorithm~\ref{alg:skipLists:constructWholeMembershipProof}) and a
whole advancement proof
(Algorithm~\ref{alg:skipLists:constructWholeAdvancementProof}) are
constructed.

\begin{algorithm}
\caption[Construction of a membership proof within an
AASL.]{\small\algsig{alg:skipLists:constructWholeMembershipProof}{ConstructMembershipProof}
$(i, n) \Rightarrow E$.  Return a membership proof $E$ for the $i$-th
element of an AASL, verifiable against the $n$-th authenticator, where
$n \geq i$.}
\begin{algorithmic}[1]
\small

\STATE $E \leftarrow \varnothing$ \COMMENT{The proof.}

\STATE $j \leftarrow i$ \COMMENT{Current element.}

\REPEAT

\STATE $C \leftarrow$ \algref{alg:skipLists:singleElementComponent}
$(j)$

\STATE $E \leftarrow E \| C$

\STATE $l \leftarrow$ \algref{alg:skipLists:singleHopTraversalLevel}
$(j, n)$

\STATE $j \leftarrow j + 2^l$

\UNTIL{$j > n$}

\STATE Return $E$

\end{algorithmic}
\end{algorithm}

\begin{algorithm}
\caption[Construction of an advancement proof within an
AASL.]{\small\algsig{alg:skipLists:constructWholeAdvancementProof}{ConstructAdvancementProof}
$(i, n) \Rightarrow A$.  Construct an advancement proof from the $i$-th
authenticator of an AASL to the $n$-th authenticator, where $n > i$.}
\begin{algorithmic}[1]
\small

\STATE $A \leftarrow \varnothing$ \COMMENT{The proof.}

\STATE $j \leftarrow i$ \COMMENT{Current element.}

\WHILE{$j < n$}

\STATE $l \leftarrow$ \algref{alg:skipLists:singleHopTraversalLevel}
$(j, n)$

\STATE $j \leftarrow j + 2^l$

\STATE $C \leftarrow$ \algref{alg:skipLists:singleElementComponent}
$(j)$

\STATE $A \leftarrow A \| C$

\ENDWHILE

\STATE Return $A$

\end{algorithmic}
\end{algorithm}

\section{Proofs of Additional Claims}

In this appendix, we prove the intuitive claims we have used in the
security analysis of the paper.

First, we prove a claim necessary for the collision-resistance theorem
(Theorem~\ref{thm:skipLists:proofCollisionResistance}), showing that
\algref{alg:skipLists:processProofComponent} is collision-resistant.

\begin{lemma}[Different proof components cannot yield the same
authenticator]
\label{lem:skipLists:differentProofComponentsSameT}

Consider two independent invocations of
\algref{alg:skipLists:processProofComponent} with inputs $(j, C)$ and
$(j', C')$ respectively.  If the two invocations yield the same result
$T$, then the inputs must be identical ($j = j'$ and $C = C'$).

\end{lemma}

In both invocations, Line~\ref{alg:skipLists:processProofComponent:T}
must yield the same result $T$.  Since $h$ is collision resistant, the
input $P$ to the hash function must be the same across invocations.

Input $P$ is constructed in the loop of
Lines~\ref{alg:skipLists:processProofComponent:loopStart} -
\ref{alg:skipLists:processProofComponent:loopEnd}, by concatenating a
hash result, produced in
Line~\ref{alg:skipLists:processProofComponent:L}, to the running $P$ in
every iteration.  To ensure that $P$ is the same in both invocations,
the loop must be iterated the same number of times (so as to construct
$P$'s of the same bit length), and all appended $L$-elements in the
respective invocations must be identical.

At every iteration of the loop,
Line~\ref{alg:skipLists:processProofComponent:L} computes the current
$L$ by hashing together the index $j$ of the assumed AASL element to
which the current proof component should correspond, the iteration
number $l$ (which is, by default, the same across invocations), the
purported data value of the $j$-th AASL element, and the $l$-th
authenticator value contained in the proof component.  Again, due to the
collision resistance of the hash function $h$, the $L$ values computed
in the two invocations can be identical only if the input index $j$ is
equal across invocations, and, similarly, if all parts of the input
proof component $C$ are respectively identical.  This means that two
invocations of \algref{alg:skipLists:processProofComponent} for inputs
$(j, C)$ and $(j', C')$ can yield the same $T$ if and only if $j = j'$
and $C = C'$. \qed

As mentioned in the paper, second pre-image resistance is a corollary of
the collision resistance theorem.

\begin{theorem}[AASL Membership Proof Second Pre-image Resistance]
\label{thm:skipLists:proofSecondPreImageResistance}

Consider a membership proof $E^{i,n,d}$ that verifies against
authenticator $T$ the membership claim $\langle i, n, d \rangle$, where
$0 < i \leq n$, and $d$ is a data value.  A computationally bound
adversary cannot construct efficiently a different membership proof $E'$
verifiable against the same authenticator $T$ that authenticates a
conflicting membership claim.

\end{theorem}

Suppose that an adversary can, in fact, construct efficiently such a
second proof $E'$ for the membership claim $t' = \langle i, n', d'
\rangle$, where $n \neq n'$ or $d \neq d'$.  This means that he has an
efficient way to construct collisions as well: he creates a legitimate
AASL, picks a random position and constructs a membership proof $E$ for
it, then constructs another membership proof $E'$ for a different data
element in the same position.  The two proofs would be a collision as
defined in Theorem~\ref{thm:skipLists:proofCollisionResistance}.
However, we have already shown that collisions are not possible, so the
proof machinery must also be second pre-image resistant. \qed

Before we can prove the authenticator agreement claims, we must first
establish that skip list traversal, as described by
\algref{alg:skipLists:singleHopTraversalLevel}, follows the rules of the
skip list, specifically that both source and destination of an $l$-level
hop are divisible by $2^l$.

\begin{lemma}[Correctness of skip list traversal]
\label{lem:skipLists:correctTraversal}
In both advancement paths and membership proof paths, as accepted by the
verification algorithms
\algref{alg:skipLists:processWholeAdvancementProof} and
\algref{alg:skipLists:processWholeMembershipProof}, respectively, every
hop from element $i$ to element $j$ has length $2^l$, such that $2^l$
divides both $i$ and $j$.
\end{lemma}

We prove this claim informally, by inspection of the corresponding
algorithms.

The path of an advancement is verified by
\algref{alg:skipLists:processWholeAdvancementProof}.  The verified path
starts with the source element $i$, given in the input parameters to the
algorithm, and proceeds by increments of $2^l$ in
Line~\ref{alg:skipLists:processWholeAdvancementProof:increment} inside
the loop.  The exponent $l$ of the path length is determined by
\algref{alg:skipLists:singleHopTraversalLevel}, given the current
element $j$ and the ultimate destination $n$ of the advancement.

Similarly, a membership proof path is verified by
\algref{alg:skipLists:processWholeMembershipProof}.  The path starts
with the source element $i$ where the element to be authenticated is
claimed to reside in the input parameters.  Then the path proceeds by
increments of $2^l$ in
Line~\ref{alg:skipLists:processWholeMembershipProof:firstIncrement} for
the first hop and
Line~\ref{alg:skipLists:processWholeMembershipProof:increment} for all
subsequent hops.  Both lines receive their $l$ from the result of
\algref{alg:skipLists:singleHopTraversalLevel}, given the current
element $j$ ($i$ in the case of
Line~\ref{alg:skipLists:processWholeMembershipProof:firstIncrement}) and
the ultimate destination $n$ of the membership proof.

For both types of paths, it suffices to show that the $l$ computed by
\algref{alg:skipLists:singleHopTraversalLevel} is such that $2^l$
divides $j$.  Then it must also divide the destination $j + 2^l$.
\algref{alg:skipLists:singleHopTraversalLevel} uses as a fall-through
selection of $l$ the value $0$, which is consistent with the claim,
since $2^0 = 1$ divides all elements.  When the loop in the algorithm is
executed at least once, the variable $L$ returned is always one that has
passed the conditional check of the loop, that is, $2^L$ divides the
source element $j$ (called $i$ in
\algref{alg:skipLists:singleHopTraversalLevel}).

We have shown that membership proof paths, and paths of single
advancements satisfy the claim.  For advancement paths of multiple
advancements the claim also holds, since connected advancements share an
element: the earlier one ends where the later one begins.  This means
there are no additional hops in the resulting advancement path to those
included in the individual advancements, which already satisfy the claim
as we showed above. \qed

We continue by analyzing the concept of the basis.  We use the two
lemmata below in authenticator agreement.

\begin{lemma}[Correspondence of bases to binary representations]
\label{lem:skipLists:binaryBasis}
Given the $l$-th AASL element, if the binary representation $b_k b_{k-1}
\ldots b_{0}$ of $l$ has a $0$ in bit position $i$, then the
corresponding basis vector $B^l$ has an empty value in vector position
$i$, and a non-empty value otherwise.
\end{lemma}
Bases are changed only via
\algref{alg:skipLists:processAdvancementProofComponent}, so we
concentrate on that to prove this lemma.  We prove the lemma by
induction on all AASL elements, and for every element on all hop lengths
leading to that element.

By definition, the base case holds, since $B^0$ has only empty values,
just as the binary representation of $0$ has only $0$'s.

We assume that the lemma holds for all bases up to that of element
$k-1$: that is, the basis vector for AASL element index $m \leq k - 1$
has an empty value in position $l$ if and only if the binary
representation of $m$ has a 0 in bit position $l$.  We show that this
must also hold for the basis $B^k$ that corresponds to element index
$k$.

\algref{alg:skipLists:processAdvancementProofComponent} yields the basis
$B^k$ for element index $k$ whenever its input contains the source
element index $j$ and the hop level $l$ and $j = k - 2^l$.
\algref{alg:skipLists:processWholeAdvancementProof} expects the outcome
of such an invocation to be $B^k = B^{j + 2^l}$ in
Line~\ref{alg:skipLists:processWholeAdvancementProof:B}.

There are $f_k + 1$ ways in which
\algref{alg:skipLists:processAdvancementProofComponent} can be invoked
to return $B^k$, one for each different level $l$ at which an
advancement path reaches element $k$.  This is because, as shown in
Lemma~\ref{lem:skipLists:correctTraversal},
Line~\ref{alg:skipLists:processWholeAdvancementProof:l} of
\algref{alg:skipLists:processWholeAdvancementProof} can only return $l$s
such that the source (and consequently the destination) of the level-$l$
hop (computed in
Line~\ref{alg:skipLists:processWholeAdvancementProof:increment}) is
divisible by $2^l$.  Since $f_k$ is the exponent of the largest power of
2 that divides $k$, as per Equation~\ref{eqn:skipLists:f}, there are
$f_k + 1$ invocations of
Line~\ref{alg:skipLists:processWholeAdvancementProof:B} that make
variable $j$ in
Line~\ref{alg:skipLists:processWholeAdvancementProof:increment} to take
the value $k$.

We consider invocations of
\algref{alg:skipLists:processAdvancementProofComponent} for all $l$ such
that $0 \leq l \leq f_k$, where $j = k - 2^l$ and $B = B^j$.  All of the
possible input bases $B = B^j$ correspond to element indices $j$ that
precede $k$, and as a result are covered by the inductive hypothesis,
above.

In the ``then'' branch of the conditional
(Line~\ref{alg:skipLists:processAdvancementProofComponent:or}), the
previous AASL element index $j$ had a 0 in the $l$-th bit position of
its binary representation.  By turning that 0 to a 1 via assigning a
non-empty value to the $l$-th basis vector element, we add $2^l$ to the
binary representation of $j = k - 2^l$, and we therefore reach the
binary representation for $k$.

If, instead, the ``else'' branch of the conditional is executed, the
previous basis vector must have had a non-empty value in its $l$-th
position, and, as a result, the binary representation of $j$ must have
had a 1 in the $l$-th bit position of its binary representation.  The
algorithm places empty values in all vector positions from the $l$-th
one upwards that contain non-empty values and sets to a non-empty value
(the value of the $\mathit{carry}$ variable) the first vector position
$m > l$ that it finds containing an empty value.  This translates into
zeroing out all 1 bits in the binary representation of $j$ from the
$l$-th to the $m-1$-st bit positions, and placing a 1 in the formerly 0
$m$-th bit.  Zeroing out a 1 bit in position $p$ means subtraction by
$2^p$, so the result of the operation is to add $(2^m - \sum_{p = l}^{m
- 1}{2^p} = 2^l)$ to the binary representation of $j = k - 2^l$, which
again yields the binary representation of $k$.

This proves the inductive step, and as a result the lemma holds for all
bases computed by
\algref{alg:skipLists:processAdvancementProofComponent}.\qed

\begin{lemma}[Survival of authenticators in a basis]
\label{lem:skipLists:authenticatorSurvival}
Consider a portion of an advancement path that goes through elements $e$
and $e' = e + 2^l$, for non-negative integers $e$ and $l$.  If $T^{e}$
is the authenticator for element $e$ computed by the advancement
processing algorithm after reaching that element, then the value for
$T^{e}$ is preserved by the algorithm in the basis, and still regarded
as that of $T^{e}$ during processing of element $e'$.
\end{lemma}

Informally, this lemma claims that while processing intermediate hops
between two elements that are successive multiples of $2^l$, the
advancement verification algorithm remembers the authenticator of the
first multiple, and uses its value to check the correctness of the
processed advancement component when it reaches the second multiple.

Since both $e$ and $e'$ are divisible by $2^l$, then in the binary
representation of $e$, bits $0$ through at least $l - 1$ are all 0.
Because of Lemma~\ref{lem:skipLists:binaryBasis}, all
basis elements in positions $0$ through at least $l - 1$ must be the
empty value.

\begin{subparagraph}{Case 1: $e$ and $e'$ are consecutive elements in
the advancement path.}

The advancement path takes a single hop at level $l$ from $e$ to $e'$.
To process this advancement hop, the verifier executes an iteration of
the loop in \algref{alg:skipLists:processWholeAdvancementProof} where
the local variable $j$ is equal to $e$ and the level returned in
Line~\ref{alg:skipLists:processWholeAdvancementProof:l} is $l$.

The value for $T^e$ was either passed as input $T_\mathit{prev}$ to the
algorithm, if this hop is the first in its advancement, or computed and
stored in $T_\mathit{cur}$ in the previous iteration of the loop in
Line~\ref{alg:skipLists:processWholeAdvancementProof:T}, and then copied
to $T_\mathit{prev}$ in
Line~\ref{alg:skipLists:processWholeAdvancementProof:previousCurrent}.

Trivially, therefore, the value of $T_\mathit{prev}$, which the
algorithm regards as $T^e$ during the loop iteration that starts with $j
= e$, is passed as input to
\algref{alg:skipLists:processAdvancementProofComponent} in
Line~\ref{alg:skipLists:processWholeAdvancementProof:B} and checked for
consistency in
Line~\ref{alg:skipLists:processWholeAdvancementProof:continuity}.  This
proves the claim for this case.

\end{subparagraph}

\begin{subparagraph}{Case 2: $e$ and $e'$ are not consecutive elements
in the advancement path.}

Leaving element $e$, the advancement path takes a hop at level $p$,
where $p < l$.  Therefore, during the invocation of
\algref{alg:skipLists:processAdvancementProofComponent} that takes as
input the basis of element $e$,
Line~\ref{alg:skipLists:processAdvancementProofComponent:or} is
executed.  What the algorithm regards at the time as $T^{e}$ (passed to
it in its input parameters) is placed in the $p$-th position of the
basis.  Since all vector positions up to position $l-1$ contained the
empty value before this modification, $T^{e}$ is the last (indeed, the
only) non-empty value in the newly created basis vector in positions 0
through $l-1$.

In what remains of the advancement path to $e'$, the value for $T^{e}$
is always the last non-empty element in vector positions 0 through
$l-1$.  This is the case right after advancement element $e$ has been
processed, as shown above.  We use this fact as the base case of an
inductive argument.

Assume that $T^e$ is the last non-empty value in the first $l$ elements
of the basis vector, and it occupies position $q < l$.  From
Lemma~\ref{lem:skipLists:binaryBasis}, the current element index is only
divisible, at most, by powers of 2 up to $2^{q}$.  This means that the
next advancement hop, as determined by
\algref{alg:skipLists:singleHopTraversalLevel} in
Line~\ref{alg:skipLists:processWholeAdvancementProof:l} of
\algref{alg:skipLists:processWholeAdvancementProof} can only proceed by
a hop of length that is a power of 2 up to $2^{q}$.  This only changes
the $q+1$ least significant bits of the element's binary representation.
Therefore, even if the ``else'' branch of the conditional in
\algref{alg:skipLists:processAdvancementProofComponent} is executed, the
value of $T^{e}$ is the last non-empty value before the $l$-th element
of the basis, and as a result is pushed to a higher element position of
the basis.

The only advancement hop that can eliminate $T^{e}$ from the first $l$
elements of the basis is the last one, leading to $e'$.  Then value
$T^e$ occupies position $(l-1)$ of the basis: we showed above there
cannot be any non-empty values between itself and position $l$, and the
value must be eliminated from the first $l$ positions of the basis,
since $e'$ is divisible by $2^l$ and has no 1's in its binary
representation up to and including bit position $l-1$.

This means that when element $e'$ is reached by the advancement proof
verification algorithm, the value for $T^e$ is in the basis, in position
$l-1$.  This is the basis vector position in which the algorithm expects
to find the value for $T^e$ during consistency checking in
Line~\ref{alg:skipLists:processAdvancementProofComponent:consistency} of
\algref{alg:skipLists:processAdvancementProofComponent}.  Indeed, the
last advancement proof component that is processed is the one
corresponding to element index $e'$, which has in the $l$-th position
among its included authenticators what the prover sent as $T^{e' - 2^l}
= T^{e}$.  Note that when
Line~\ref{alg:skipLists:processAdvancementProofComponent:consistency} is
executed and eliminates value $T^e$ from basis vector position $l-1$,
the local loop variable $c$ is equal to $l-1$.
\end{subparagraph}

Consequently, we have shown that the claim holds for both possible cases
of advancement paths between $e$ and $e'$, which proves the lemma.  \qed

The proofs for evolutionary collision resistance and for the
authenticator agreement lemmata rely heavily on common elements in
membership or advancement proof paths.  We proceed with two lemmata that
examine the arrangement of common elements of parallel paths.  First, we
look at common elements of parallel paths, regardless of the path type
(advancement or membership proof).  Then, we show that between two
common elements in a membership proof and advancement path, the
advancement path always takes shorter hops than the membership proof
path.

\begin{lemma}[Common elements of parallel paths]
\label{lem:skipLists:parallelPaths}
Let $i$ and $j$ be positive integers, such that $i < j$.
\begin{enumerate}
\item Consider a path $A$ that includes element $i$ and continues to $j$
or past it. There is at least one element in $[i, j]$ that is shared by
$A$ and every path that starts at or before $i$ and includes element
$j$.  The last element on path $A$ before $j$ (or $j$ if it is in path
$A$) is such an element.
\item (The mirror case) Consider a path $A$ that starts at or before
element $i$ and includes element $j$. There is at least one element in
$[i, j]$ that is shared by $A$ and every path that includes element $i$
and continues to element $j$ or past it.  The first element on $A$ after
$i$ (or $i$ if it is in path $A$) is such an element.
\end{enumerate}
\end{lemma}

We prove only the first part of the lemma.  The proof for the second
part of the lemma is a trivial ``mirror image'' of the proof for the
first part.

If path $A$ contains element $j$, then we are trivially done.

\begin{figure}
\centerline{\includegraphics{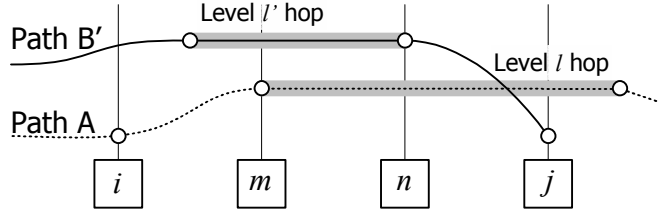}}
\caption[Two parallel, interleaved paths.]{Two parallel, interleaved
paths $A$ and $B$.  $A$ contains $i$, but not necessarily $j$.  $B$
contains $j$ but not necessarily $i$.  The thick gray lines represent
single hops, as picked by
\algref{alg:skipLists:singleHopTraversalLevel}.}
\label{fig:skipLists:ParallelPaths}
\end{figure}

Now, assume that path $A$ does not contain element $j$, and consider
Figure~\ref{fig:skipLists:ParallelPaths}.  Path $A$ must be able to
overshoot element $j$ on its way from $i$ to beyond $j$.  For this to
happen, path $A$ must advance from its last element $m$ before $j$
(i.e., $i \leq m < j$) past $j$, by a hop of level $l$ and length $2^l$,
where $2^l$ divides $m$, and $j$ must not participate in any linked list
at level $l$ or higher (i.e., $2^l$ does not divide $j$).  The end point
of this hop is $m + 2^l > j$.

Suppose $m$ is not the single common element among path $A$ and every
path that starts at or before $i$ and includes element $j$.  Then there
must be a path $B'$ that manages to overshoot element $m$ on its way to
$j$.  For this to happen, path $B'$ must advance to its first element
$n$ after $m$ (i.e., $m < n \leq j$) past $m$, by a hop of level $l'$
and length $2^{l'}$, and element $m$ must not participate in any linked
list at level $l'$ or higher (i.e., $2^{l'}$ does not divide $m$).  This
means that $l' > l$, since $m$ is divisible by $2^l$.  If $n$
participates in the linked list at level $l'$, it must be divisible by
$2^{l'}$ and, as a result, also by $2^l$.  However, that is impossible,
since $m < n \leq j < m + 2^l$.

Therefore, every path $B$ that starts at or before $i$ and includes $j$
must include element $m$, which also belongs to path $A$.\qed

\begin{lemma}[Common elements of proof and advancement paths]
\label{lem:skipLists:commonProofAdvancementPoints}
If a membership proof path has two common elements $e$ and $e'$ with an
advancement path, then every element in the proof path between $e$ and
$e'$ is also shared by that advancement path.
\end{lemma}

Consider a membership proof path and an advancement path that share
elements $e$ and $e > e'$, but share no other elements between them.

\begin{figure}
\centerline{\includegraphics{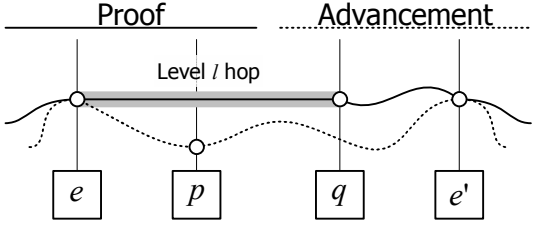}}
\caption[Proof and advancement paths.]{Proof and advancement paths
between their common elements $e$ and $e'$.  The thick gray line
indicates a single hop, as calculated by
\algref{alg:skipLists:singleHopTraversalLevel}.}
\label{fig:skipLists:ProofAndAdvancement}
\end{figure}

To prove the lemma, we suppose that none of the proof elements between
$e$ and $e'$ belong to the advancement path (see
Figure~\ref{fig:skipLists:ProofAndAdvancement}).  We show below that
this hypothesis leads to a contradiction.

After common element $e$, the two paths diverge.  The membership proof
takes a hop at level $l$, whereas the advancement takes a hop at a lower
level $l' < l$.  The advancement cannot take a hop at a level higher
than that of the proof; if such a hop were available that did not
overshoot $e'$, then the proof would have also taken it (see
\algref{alg:skipLists:singleHopTraversalLevel}).  Furthermore, the
advancement cannot take a hop at the same level $l$ as the proof,
because that would make the two paths identical between $e$ and $e'$,
which contradicts the hypothesis that intermediate membership proof
elements do not belong to the advancement path.  We call the next
element on the advancement path $p = e + 2^{l'}$, and the next element
on the membership proof path $q = e + 2^l$.

Because of Lemma~\ref{lem:skipLists:parallelPaths}, there must be a
common element between the two paths in $[p, q]$.  However, this
contradicts the hypothesis that the two paths share no elements between
$e$ and $e'$.  As a result, all membership proof elements between $e$
and $e'$ must also belong to the advancement path. \qed

Finally, we prove the two authenticator agreement lemmata.

\begin{lemma}[Authenticator agreement between a membership proof and an advancement
proof verification]
\label{lem:skipLists:authenticatorAgreementProofAdvancement}
If the membership proof verification
Algorithm~\ref{alg:skipLists:processWholeMembershipProof} and the
advancement verification Algorithm
\ref{alg:skipLists:processWholeAdvancementProof}, given an advancement
sequence and a membership proof, respectively, agree on the value of
authenticator $T^n$ for element $n$ during their independent executions,
then they also agree on the authenticator value $T^e$ of every other
earlier element $e$ ($e < n$) that the advancement path and membership
proof path have in common.
\end{lemma}

Let $k$ be the number of common elements in the two paths up to element
$n$, and $n = e_1 > e_2 > \ldots > e_k$ the common elements, from last
to first.  We prove the lemma using induction on the common elements
$e_i$, by following backwards
Algorithms~\ref{alg:skipLists:processWholeMembershipProof} and
\ref{alg:skipLists:processWholeAdvancementProof}.

The base case for $e_1$ holds from the lemma assumption, since $e_1 =
n$.

For the inductive step, we assume that the two algorithms agree on the
value of $T^{e_i}$.  We must show that the two algorithms also agree on
the value of $T^{e_{i+1}}$ when they process the corresponding proof
component to reach element $e_{i+1}$.

When the two algorithms process their respective proof component to
compute the common $T^{e_i}$ they use Equations~\ref{eqn:skipLists:L}
and \ref{eqn:skipLists:T}.  Specifically, they both compute
\begin{eqnarray}
T^{e_i} & = & h(\ldots \| L^{l}_{e_i} \| \ldots) \nonumber\\
        & = & h(\ldots \| \overbrace{h(e_i \| l \| d_i \| T^{e_i -
              2^l})}^{L^{l}_{e_i}} \| \ldots) \nonumber
\end{eqnarray}
by invoking \algref{alg:skipLists:processProofComponent} in
Line~\ref{alg:skipLists:processWholeMembershipProof:T} of
\algref{alg:skipLists:processWholeMembershipProof} and
Line~\ref{alg:skipLists:processWholeAdvancementProof:T} of
\algref{alg:skipLists:processWholeAdvancementProof}. Since $h$ is
collision resistant, when the two algorithms process element $e_i$ they
must agree on the values of all $T^{e_i - 2^l}$, for every level $l$ of
linked lists in which element $e_i$ participates.

Consider what happens in the two paths between elements $e_{i+1}$ and
$e_i$.  Common element $e_{i+1}$ must be the membership proof element
immediately preceding $e_i$, because of
Lemma~\ref{lem:skipLists:commonProofAdvancementPoints}.  Therefore,
because of Lemma~\ref{lem:skipLists:correctTraversal}, $e_{i} = e_{i+1}
+ 2^{l'}$ for some non-negative $l'$.  The advancement hop that arrives
at $e_i$ must be at the same level $l'$ or lower level.  This is because
a higher-level $l'' > l'$ hop would have taken the advancement path from
$e_{i+1}$ to element $e_{i+1} + 2^{l''}$, which must lie beyond $e_i =
e_{i+1} + 2^{l'}$.  Therefore, the advancement path between $e_{i+1}$
and $e_i$ follows either a single hop of level $l'$ and length $2^{l'}$,
which is identical to that followed by the membership proof path, or a
sequence of shorter hops at levels lower then $l'$.

\begin{subparagraph}{Case 1: The advancement path is identical to the
membership proof path.} The value for $T^{e_{i+1}}$ used to compute
$T^{e_i}$ in the two algorithms while processing element $e_i$ is the
same as that known by the algorithms while processing the previous
element $e_{i+1}$, from
Line~\ref{alg:skipLists:processWholeMembershipProof:continuity} of
\algref{alg:skipLists:processWholeMembershipProof} and
Line~\ref{alg:skipLists:processWholeAdvancementProof:continuity} of
\algref{alg:skipLists:processWholeAdvancementProof}, which proves the
inductive step.

\end{subparagraph}

\begin{subparagraph}{Case 2: The advancement path is not identical to
the membership proof path.} We must establish that the value of
$T^{e_{i+1}}$ that \algref{alg:skipLists:processWholeAdvancementProof}
computes while processing element $e_{i+1}$ is the same as that known
while processing the next common element $e_i$.  This follows from
Lemma~\ref{lem:skipLists:authenticatorSurvival}, since elements
$e_{i+1}$ and $e_i$ are successive multiples of $2^{l'}$.

As a result, the value for $T^{e_{i+1}}$ produced by the advancement
verification algorithm while processing element $e_{i+1}$ is the same as
the value for $T^{e_{i+1}}$ used by the membership proof verification
algorithm while processing element $e_i$.  This is the same value as
that for $T^{e_{i+1}}$ produced by the proof verification algorithm
while processing element $e_{i+1}$, as seen in
Line~\ref{alg:skipLists:processWholeMembershipProof:continuity} of
\algref{alg:skipLists:processWholeMembershipProof}.  This proves the
inductive step.
\end{subparagraph}

The inductive step holds for both possible cases of advancement paths,
and as a result, the inductive argument holds, proving the lemma.  \qed

\begin{lemma}[Authenticator agreement between two independent
advancement paths]
\label{lem:skipLists:commonAdvancementPoints}
If two invocations of the advancement verification
Algorithm~\ref{alg:skipLists:processWholeAdvancementProof}, given two
advancement sequences, respectively, agree on the value of authenticator
$T^n$ computed after reaching element $n$ during their independent
executions, then they also agree on the authenticator value $T^e$
computed after reaching every other earlier element $e$ ($e < n$) that
the advancement paths have in common.
\end{lemma}

This proof is similar in structure to that of the preceding lemma.

Let $k$ be the number of common elements in the two paths up to element
$n$, and $n = e_1 > e_2 > \ldots > e_k$ the actual elements, from last
to first.  We prove the lemma using induction on the common elements
$e_i$, by following backwards two invocations of
Algorithm~\ref{alg:skipLists:processWholeAdvancementProof}.

The base case for $e_1$ holds from the lemma assumption, since $e_1 =
n$.

For the inductive step, we assume that the two algorithms agree on the
value of $T^{e_i}$, after reaching element $e_i$.  We must show that the
two algorithms also agree on the value of $T^{e_{i+1}}$ after they reach
element $e_{i+1}$.

When the two algorithms process their respective proof component to
compute the common $T^{e_i}$ they use Equations~\ref{eqn:skipLists:L}
and \ref{eqn:skipLists:T}.  Specifically, they both compute
\begin{eqnarray}
T^{e_i} & = & h(\ldots \| L^{l}_{e_i} \| \ldots) \nonumber\\
        & = & h(\ldots \| \overbrace{h(e_i \| l \| d_i \| T^{e_i -
              2^l})}^{L^{l}_{e_i}} \| \ldots) \nonumber
\end{eqnarray}
by invoking \algref{alg:skipLists:processProofComponent} in
Line~\ref{alg:skipLists:processWholeAdvancementProof:T} of
\algref{alg:skipLists:processWholeAdvancementProof}.  Since $h$ is
collision resistant, when the two algorithm runs process element $e_i$
they must agree on the values of all $T^{e_i - 2^l}$, for every level
$l$ of linked lists in which element $e_i$ participates.

Consider what happens in the two paths between elements $e_{i+1}$ and
$e_i$.

\begin{subparagraph}{Case 1: Element $e_{i+1}$ immediately precedes
element $e_i$ in both paths.}  Both paths advance from $e_{i+1}$ to
$e_{i}$ in a single hop at level $l$, of length $2^l$.

As shown above, the two runs agree on the value of $T^{e_i - 2^l}$.
Since $e_i - 2^l = e_{i+1}$, and from
Line~\ref{alg:skipLists:processWholeAdvancementProof:continuity} of
\algref{alg:skipLists:processWholeAdvancementProof}, the value for
$T^{e_{i + 1}}$ while processing element $e_i$ must be identical to the
value that the two runs compute for $T^{e_{i+1}}$ after processing the
advancement at the previous element $e_{i + 1}$.
\end{subparagraph}

\begin{subparagraph}{Case 2: Element $e_{i+1}$ does not immediately
precede element $e_i$ in at least one of the paths.}  The two paths
merge from two different immediate sources to element $e_i$ on their way
from element $e_{i+1}$.  Because of
Lemma~\ref{lem:skipLists:correctTraversal}, for some
non-negative integers $0 \leq l < l'$ without loss of generality, the
element immediately preceding $e_i$ on the first path is $p = e_i -
2^l$, and on the second it is $q = e_i - 2^{l'}$.  Note that $q < p$.

Lemma~\ref{lem:skipLists:parallelPaths} guarantees
that there must be a common element between the two paths in $[q, p]$,
since the first path starts at or before $q$ and reaches $p$ on its way
to $e_i$, whereas the second path starts at $q$ and goes past $p$ on its
way to $e_i$.  Since $q$ is the element immediately preceding $e_i$ on
the second path, it must be the common element that
Lemma~\ref{lem:skipLists:parallelPaths} anticipates.
Therefore, $e_{i+1} = q$.

Because of Lemma~\ref{lem:skipLists:authenticatorSurvival},
both runs of the advancement verification algorithm agree on the value
of $T^{e_{i+1}}$ after processing element $e_{i+1}$ and after reaching
element $e_i$.

\end{subparagraph}

The inductive step holds for both possible cases of advancement path
commonalities and, as a result, the inductive argument holds, proving
the lemma.  \qed

\section{\label{sec:skipLists:implementation}Implementation}

We implement authenticated append-only skip lists using Java.  We focus
here on a disk-based implementation, since it allows much larger data
sequences than any memory-only implementation can, as well as
persistence in the face of machine reboots.

An AASL is stored on disk as a linear file that consists of a
\emph{preamble} and a sequence of \emph{element entries}, one for each
element currently contained in the AASL.  An element entry consists of
a data section and an authenticator section.

The data section primarily holds the datum stored in the associated AASL
element.  This is the datum that participates in the computation of
authenticators, as per Equations~\ref{eqn:skipLists:L} and
\ref{eqn:skipLists:T}.  We call this the \emph{sensitive datum}.  Every
element in a single AASL has sensitive data of a constant length, which
is set when the AASL is initially created.

The data section of element entries may also contain an
\emph{insensitive datum}.  This is also a fixed-length bit string.
However, it does not participate in authenticator computations.
Insensitive data may be useful information to the maintainer, collocated
with the sensitive data for access efficiency, that need not be
authenticated to remote verifiers of the AASL.  Since insensitive data
do not participate in authenticator computations, they can be changed at
will by the AASL maintainer unobtrusively to AASL verifiers.

The authenticator section of an element entry contains the authenticator
computed for that element.

The preamble of the AASL file contains the lengths in bytes of the
sensitive and insensitive data in element entries, and the element
position of the last incorporated element into the AASL.

An empty AASL contains exactly one element entry: the entry for element
$0$, which is a special entry.  Element entry $0$ has inconsequential
sensitive and insensitive data.  Only its authenticator is meaningful.
This authenticator is a value from the result domain of the hash
function used, and it is agreed upon among all users of the AASL in
advance.

Our implementation has a deviation from the abstract design of AASLs
described in Section~\ref{sec:skipLists:design}.  We slightly modify how
authenticators are computed for elements of odd indices, which only
participate in a single linked list.  For such elements we skip the
outer hash operation described in Equation~\ref{eqn:skipLists:T}, from
concatenated partial authenticators to the actual authenticator of the
element.  Since odd elements have only a single partial authenticator,
that single partial authenticator is sufficient to ensure the collision
resistance of AASL digests, and can serve as the actual authenticator of
the element.  Furthermore, since half of the element indices are odd,
this savings in computation can be significant compared to the overall
computation required by AASL operations.

Another implementation optimization in the implemented AASLs deals with
authenticator redundancy in membership and advancement proofs.  In the
idealized algorithms \algref{alg:skipLists:processWholeMembershipProof}
and \algref{alg:skipLists:processWholeAdvancementProof}, authenticators
computed for the previous proof component are compared against the
corresponding authenticator included in the next proof component (see
Lines~\ref{alg:skipLists:processWholeMembershipProof:continuity} and
\ref{alg:skipLists:processWholeAdvancementProof:continuity},
respectively).  Since we compute these authenticators in the process of
verifying membership and advancement proofs anyway, there is no need
also to include them in the proofs themselves.  Consequently, we skip
such authenticators in the AASL implementation.

{
\tiny
\begin{verbatim}
$Id: PODC2003.tex,v 1.23 2003/02/07 01:11:25 maniatis Exp $
\end{verbatim}
}

\end{document}